\documentclass[acmlarge,screen]{acmart}
\AtBeginDocument{%
  \providecommand\BibTeX{{%
    \normalfont B\kern-0.5em{\scshape i\kern-0.25em b}\kern-0.8em\TeX}}}

\setcopyright{none}

\renewcommand\footnotetextcopyrightpermission[1]{} % removes footnote with conference information in first column
\usepackage{multirow}
\usepackage{subfig}
\usepackage{subcaption}

\usepackage{utfsym}

\begin{document}

%%
%% The "title" command has an optional parameter,
%% allowing the author to define a "short title" to be used in page headers.

\title{HAIGEN: Towards Human-AI Collaboration for Facilitating Creativity and Style Generation in Fashion Design}

\author{\href{https://orcid.org/0000-0003-3029-8146}{Jianan Jiang}, \href{https://orcid.org/0000-0001-8697-1817}{Di Wu}, \href{https://orcid.org/0000-0003-2091-5163}{Hanhui Deng}}
\affiliation{
  \institution{Hunan University}
  \city{Changsha}
  \state{Hunan}
  \country{China}}

\author{\href{https://orcid.org/0009-0001-7607-7106}{Yidan Long}, \href{https://orcid.org/0009-0007-4530-2589}{Wenyi Tang}, \href{https://orcid.org/0009-0005-8107-4590}{Xiang Li}}
\affiliation{
  \institution{Hunan Normal University}
  \city{Changsha}
  \state{Hunan}
  \country{China}}

\author{\href{https://orcid.org/0000-0003-3267-3317}{Can Liu}}
\affiliation{
  \institution{City University of Hong Kong}
  \city{}
  \state{Hong Kong}
  \country{China}}

\author{\href{https://orcid.org/0000-0002-3020-3736}{Zhanpeng Jin}}
\affiliation{
  \institution{South China University of Technology}
  \city{Guangzhou}
  \state{Guangdong}
  \country{China}}

\author{\href{https://orcid.org/0009-0007-9465-4273}{Wenlei Zhang}, \href{https://orcid.org/0009-0007-9746-1557}{Tangquan Qi}}
\affiliation{
  \institution{Wondershare Technology}
  \city{Changsha}
  \state{Hunan}
  \country{China}}

\authorsaddresses{D. Wu is the corresponding author (Email: dwu@hnu.edu.cn).}

\renewcommand{\shortauthors}{Jiang et al.}

\begin{abstract}
The process of fashion design usually involves sketching, refining, and coloring, with designers drawing inspiration from various images to fuel their creative endeavors. However, conventional image search methods often yield irrelevant results, impeding the design process. Moreover, creating and coloring sketches can be time-consuming and demanding, acting as a bottleneck in the design workflow. In this work, we introduce HAIGEN (\textbf{H}uman-\textbf{AI} Collaboration for \textbf{GEN}eration), an efficient fashion design system for Human-AI collaboration developed to aid designers. Specifically, HAIGEN consists of four modules. T2IM, located in the cloud, generates reference inspiration images directly from text prompts. With three other modules situated locally, the I2SM batch generates the image material library into a certain designer-style sketch material library. The SRM recommends similar sketches in the generated library to designers for further refinement, and the STM colors the refined sketch according to the styles of inspiration images. Through our system, any designer can perform local personalized fine-tuning and leverage the powerful generation capabilities of large models in the cloud, streamlining the entire design development process. Given that our approach integrates both cloud and local model deployment schemes, it effectively safeguards design privacy by avoiding the need to upload personalized data from local designers. We validated the effectiveness of each module through extensive qualitative and quantitative experiments. User surveys also confirmed that HAIGEN offers significant advantages in design efficiency, positioning it as a new generation of aid-tool for designers.~\footnote{Accepted to Proceedings of the ACM on Interactive, Mobile, Wearable and Ubiquitous Technologies (ACM IMWUT/UbiComp 2024). DOI:~\url{https://dl.acm.org/doi/10.1145/3678518}.}
\end{abstract}

\begin{CCSXML}
<ccs2012>
    <concept>
        <concept_id>10003120.10003138.10003140</concept_id>
        <concept_desc>Human-centered computing~Ubiquitous and mobile computing</concept_desc>
        <concept_significance>500</concept_significance>
        </concept>
   <concept>
       <concept_id>10010147.10010178</concept_id>
       <concept_desc>Computing methodologies~Artificial intelligence</concept_desc>
       <concept_significance>500</concept_significance>
       </concept>
   <concept>
       <concept_id>10010405.10010469</concept_id>
       <concept_desc>Applied computing~Arts and humanities</concept_desc>
       <concept_significance>300</concept_significance>
       </concept>
 </ccs2012>
\end{CCSXML}

\ccsdesc[500]{Human-centered computing~Ubiquitous and mobile computing}
\ccsdesc[500]{Computing methodologies~Artificial intelligence}
\ccsdesc[300]{Applied computing~Arts and humanities}

\keywords{Human-AI Collaboration; Generative Artificial Intelligence; Personalized Fashion Design}

\maketitle

%%%%%%%%%%%%%%%%%%%%%%%%%%%%%% INTRODUCTION %%%%%%%%%%%%%%%%%%%%%%%%%%%%%%
\section{INTRODUCTION}
In the traditional fashion design process, when designers conceive new ideas, they often begin by using sketches to roughly outline the overall structure of their creations. Subsequently, they search for relevant sources of material inspiration, often turning to various online platforms, design publications, and other channels to refine and iterate upon their sketch designs. Notably, the initial sketching stage can be challenging, especially for junior designers. Accessing pre-existing sketch templates for customization and modification would thus serve as a valuable resource for many designers. Following the sketching stage, designers often seek inspiration to breathe life into their creations through a coloring process, enhancing the depth and realism of the sketches. However, this stage is laborious and time-consuming. Therefore, a quick sketch coloring tool would enable them to preview the sketch coloring effect in advance and speed up their design process. Throughout the entire design process, searching for design inspiration is crucial~\cite{jonson2005design}. However, designers often encounter difficulties due to the limited results provided by many search engine functions. This frequently results in designers forgetting key details of their original inspiration, significantly impacting their overall design efficiency.

\begin{figure}[tb]
\centering
\includegraphics[width=\linewidth]{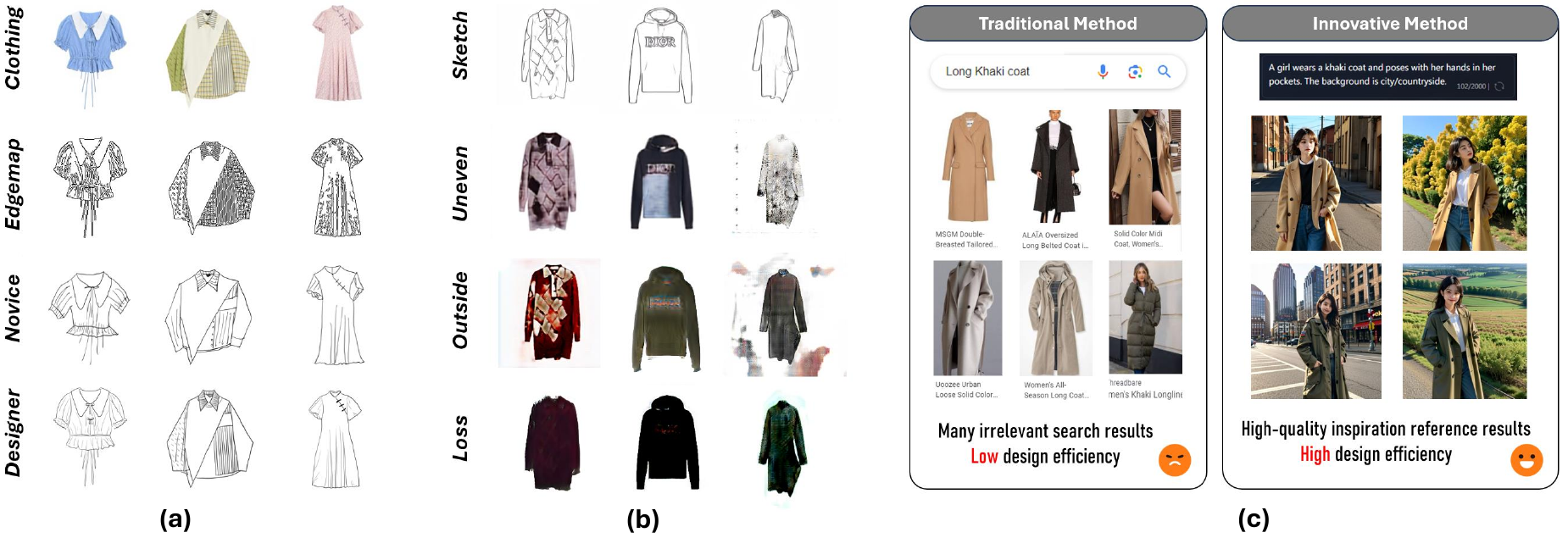}
\caption{(a) Three distinct methods of deriving sketches from a provided clothing image are presented, highlighting noticeable stylistic differences among them. (b) Highlights three issues that emerge during the sketch coloring process, influencing the designer's overall creative workflow. (c) Our approach distinguishes itself from previous methods used for searching inspirations. With our method, it can generate images that closely align with the designer's inner thoughts based on detailed text descriptions.}
\vspace{-3mm}
\label{fig:sample}
\end{figure}

Recent years have witnessed substantial advancements in Generative Artificial Intelligence (AI), particularly with the introduction of Generative Adversarial Networks (GAN)~\cite{goodfellow2014generative} and Diffusion Models (DM)~\cite{ho2020denoising}. These innovations have revolutionized the field of image generation by enabling the creation of highly realistic, diverse, and novel objects. This transformation is reshaping the landscape of the design field, leading an increasing number of designers to incorporate artificial intelligence tools into their creative processes.

In the sketching stage, most methods are limited to extracting the outline and partial details of objects~\cite{canny1986computational, xie2015holistically, wang2017bayesian, zhu2021sketch, bhunia2022doodleformer}, making it challenging to accurately represent complex details with reasonable sketch strokes. There are obvious stylistic differences in the sketching strokes of different people. Fig.~\ref{fig:sample} (a) shows the differences between synthetic edge maps~\cite{canny1986computational}, sketches without professional knowledge, and sketches with professional knowledge. In the sketch coloring stage, most research~\cite{liu2021self, wu2023styleme, zhang2023unified} relies on GAN to achieve style transfer from sketches to images. However, these methods often lead to problems such as uneven color distribution, coloring outside the sketch outline, and loss of sketch stroke details, as shown in Fig.~\ref{fig:sample} (b). These problems make it difficult for designers to achieve a perfect preview of sketch shading effects, thereby reducing design efficiency and affecting the overall design experience.

In the most critical stage of inspiration search in the entire creation, traditional methods often require designers to use as concise descriptors as possible to search for relevant materials. This is very challenging for people with weak language skills and professional abilities. \cite{schultheiss2023misplaced} demonstrates that users with limited familiarity with search engines tend to depend more on traditional methods like Google, compared to those with greater expertise. This reliance increases the likelihood of obtaining inaccurate results. In recent years, tools such as DALL-E~\cite{ramesh2021zero} and Midjourney~\cite{midjourney} have emerged, allowing designers to directly enter comprehensive descriptions of their ideas to generate inspiration materials that are as consistent as possible. Fig.~\ref{fig:sample} (c) provides a visual comparison between these two inspiration search methods. However, these tools impose significant hardware requirements making it difficult for designers to customize the model in personalized styles. Additionally, their openness raises concerns about potential privacy breaches, particularly regarding designers’ confidential inspirations.

To address these challenges, we present the HAIGEN (\textbf{H}uman-\textbf{AI} Collaboration for \textbf{GEN}eration), a Human-AI collaboration system aimed at enhancing the creative efficiency of fashion designers. It comprises the \textit{Text-to-Image Cloud Module} (T2IM) deployed in the cloud, as well as the \textit{Image-to-Sketch Local Module} (I2SM), \textit{Sketch Recommendation Module} (SRM), and \textit{Style Transfer Module} (STM) situated locally. Designers can swiftly access desired inspiration materials by harnessing the robust cross-modal generation capabilities offered by cloud-based large models. These materials can then be seamlessly integrated with multiple local small models to facilitate the progression of the entire design process. Specifically, T2IM, based on SD 1.5~\cite{rombach2022high}, generates images from text prompts, streamlining material search and providing inspiration for designers. We employ the LoRA~\cite{hu2021lora} for fine-tuning SD and combine it with ControlNet~\cite{zhang2023adding} to further optimize the edge, structure, color, and other details of the text-to-image generation process. I2SM is a GAN-based model for converting images into sketches, featuring the APSN module to capture a designer's personalized sketch style. The DSMFF module is introduced to fuse contour and detail features within multi-scale features, thereby generating numerous designer-style sketches as initial templates. SRM, based on the ViT~\cite{dosovitskiy2020vit}, recommends similar sketch templates to designers. STM is a DDIM-based~\cite{song2020denoising} model employed for sketch coloring. It incorporates the content feature of the sketch and the style feature of the reference image into the model's diffusion process explicitly and implicitly. Utilizing our innovative CCAM, it can generate high-quality sketch coloring renderings based on the style of reference images. Notably, our approach, structured with a cloud and local separation architecture, effectively safeguards design privacy by exclusively downloading inspirations generated in the cloud without the need to upload local personalized data, distinguishing it from other existing methods~\cite{ramesh2021zero, midjourney}. 

In summary, our contributions can be summarized as follows:

\begin{itemize}
\item We propose a Human-AI collaborative auxiliary design system, allowing designers to combine cloud-based large models and local-based small models to achieve efficient design processes. All models can be personalized and fine-tuned, providing a user-friendly experience for designers to pursue personalized design. Moreover, the integration of cloud and local architecture ensures stable privacy protection for designers' creativity.
\item We introduce a comprehensive SD model deployed in the cloud as a text-to-image cross-modal solution. This innovation enables the model to rapidly learn new image styles and adjust details within the generated image. By inputting relevant text prompts, it can generate multi-style, multi-aspect ratio, and multi-resolution reference images. This feature empowers designers by efficiently providing rich design inspiration.
\item Our image-to-sketch generation model offers a distinctive approach to capturing designer style. By leveraging the innovative APSN module, it learns design style features from designers' historical data and seamlessly incorporates these style features into newly generated sketches. This capability empowers designers to create adequate sketch templates with personalized styles rapidly, thereby enhancing overall design efficiency.
\item Our local-based sketch-to-image style fusion framework represents a breakthrough approach to sketch coloring stylization. We create a supervised diffusion model by explicitly conditioning on locally refined personalized sketches and implicitly conditioning on cloud-generated reference images. The resultant image encapsulates the style of the reference image while preserving the unique design style and intricate details of the sketch.
\end{itemize}

\textbf{Note:} Our HAIGEN system possesses the following key characteristics: (i) \underline{\textbf{Distributed System}}: We introduce a cloud-local collaboration solution, harnessing the powerful generation capabilities of large cloud models to inspire design creativity. Complemented by a series of local small models, this approach facilitates the entire creative process. Designers can seamlessly integrate inspirational reference images generated in the cloud into the local development process, fostering efficient design and development. Furthermore, the solution effectively isolates local development data from the cloud, ensuring user privacy protection. (ii) \underline{\textbf{Ubiquitous Computing}}: We present a Human-AI collaboration solution, facilitating rapid personalized fine-tuning of styles and multi-style fusion for cloud models. Additionally, our solution achieves lightweight design, as well as the capture and fusion of designers' historical creative styles for local models. This enables any designer in any scenario to seamlessly integrate our system at a low cost for personalized fashion design development, thereby enhancing overall design efficiency. (iii) \underline{\textbf{Applicability to Different Fields}}: With the trend of large-scale model development shifting towards vertical field applications, and Human-AI collaborative development emerging as an inevitable trend, our model cloud-local collaboration and personalized fine-tuning concepts are versatile and can be applied to various fields. These include medical image processing in the medical field, model structure design in the architectural field, teaching plan arrangement in the education field, and more.

The subsequent sections of this paper are structured as follows: Section~\ref{sec:work} provides the related works. Section~\ref{sec:overview} encompasses a user study and our system overview. Section~\ref{sec:system} conducts a detailed exposition of each module in our system. Section~\ref{sec:experiments} and Section~\ref{sec:survey} present the experimental results and user surveys. Section~\ref{sec:conclusion} concludes our work.

%%%%%%%%%%%%%%%%%%%%%%%%%%%%%% Related Work %%%%%%%%%%%%%%%%%%%%%%%%%%%%%%
\section{Related Work}
\label{sec:work}

\subsection{Generative AI in Fashion Design}
Generative AI in fashion design could be composed of inspiration searching stage, sketching stage, and sketch coloring stage. The inspiration searching stage corresponds to the Text-to-Image Generation task. Early methods, like Srivastava \textit{et al.}~\cite{srivastava2012multimodal}, attempted to employ deep Boltzmann machines for learning the modeling of images and text. However, these approaches were notably constrained in terms of complexity and realism. With the advent of Generative Adversarial Networks (GAN)~\cite{goodfellow2014generative}, particularly Conditional GAN~\cite{mirza2014conditional}, models~\cite{reed2016generative, sarafianos2019adversarial} have been developed that can effectively capture the intricate relationship between text and images. In recent years, the introduction of models such as Stable Diffusion~\cite{rombach2022high} and CLIP~\cite{radford2021clip} has ushered in a new era for text-to-image generation. Some of these newer models~\cite{zhang2023adding, ruiz2023dreambooth, kumari2023multi} are capable of generating more finely detailed images and offer greater control over the generated visuals.

The sketching stage corresponds to the Image-to-Sketch Generation task. Conventional edge extraction algorithms like Canny~\cite{canny1986computational} and HED~\cite{xie2015holistically} are inadequate in addressing these challenges. The images generated using these methods often exhibit poor quality and lack style consistency. Ha \textit{et al.}~\cite{ha2017neural} employed recurrent neural networks, Ge \textit{et al.}~\cite{ge2020creative} introduced a part-based generative adversarial network. Bhunia \textit{et al.}~\cite{bhunia2022doodleformer} proposed a two-stage framework. However, they ignore the unique design styles of different sketchers, and some of them are tailored for smaller objects. In the case of fashion clothing data, where objects are more detailed and complicated, it becomes crucial to accurately represent the diverse design styles and design skills of different designers.

The sketch coloring stage corresponds to the Sketch-based Style Transfer task. The introduction of AdaIN~\cite{huang2017adain} offers a practical solution. Li \textit{et al.}~\cite{li2020deep} decompose the task into texture synthesis and shading enhancement. Subsequently, Liu \textit{et al.}~\cite{liu2021self} proposed a two-stage style transfer framework to enhance the images generated in the first stage. StyleMe~\cite{wu2023styleme} further elevates the style transfer quality with a deeper network and a consistent loss function. Zhang \textit{et al.}~\cite{zhang2023unified} employed a contrastive learning method to regulate the style of generated images. However, these methods still face challenges in handling complex strokes and rectifying synthetic structures to achieve realistic images. Therefore, synthesizing images that accurately depict the content of the sketch is of paramount importance, all the while ensuring a cohesive style that aligns with the reference image.

\subsection{Human-AI Collaboration}
The rapid development of Artificial Intelligence (AI) technology has seen numerous algorithms transition from research laboratories to practical, real-world applications~\cite{wang2020human}. This transformative progress has led various fields to evolve from a paradigm of individual human efforts to a collaborative framework where humans and AI work in cooperation. For instance, Prajwal \textit{et al.}~\cite{prajwal2023towards} proposed a Human-AI collaborative emotion self-report collection framework based on smartphone keyboards to sense user emotions. Sun \textit{et al.}~\cite{sun2023inspire} introduced a customizable AI chatbot to explore innovative collaborative methods of illustration art and AI through chats with illustrators. Cho \textit{et al.}~\cite{cho2023ai} deployed AI in an actual home full of sensors and interactive devices to comprehensively study the experience of integrating AI-driven technology into daily home life. Moreover, Figueiredo \textit{et al.}~\cite{figueiredo2024powered} examined the impact of AI descriptions to aid health consumers in making personal health decisions based on user data and algorithmic output. Deng \textit{et al.}~\cite{deng2024crossgai} proposed a solution for multiple people to work collaboratively with AI tools on different devices and optimize network bandwidth to reduce latency through Lyaplov. All these studies confirm the effectiveness of Human-AI collaborative solutions.

Within the realm of design, traditional design methods require designers to autonomously navigate the entire design process, resulting in inefficiencies and even superfluous efforts. For instance, in the inspiration searching stage, designers aspire to swiftly locate the references that inspire them the most. During the sketching phase, most designers prefer to directly modify relevant templates. Furthermore, in the coloring stage, designers seek the ability to swiftly preview the suitable colors in advance and subsequently make relevant adjustments. Therefore, the development of an AI tool assumes paramount importance in minimizing the time and energy designers expend on non-innovative stages. This not only streamlines the design process but also enhances the overall efficiency of design endeavors.

\subsection{Privacy Protection}
As our world becomes increasingly technologically integrated~\cite{wu2021edgelstm, wu2022automl, wu2022deepbrain, ikeda2023interactive, sailaja2023ubifix}, the generation and aggregation of substantial data, often encompassing sensitive personal information, has become ubiquitous. This proliferation has raised critical concerns regarding the privacy and security of personal data, particularly in the context of software applications, websites, and digital platforms~\cite{jin2022exploring, saisho2023sandbox}. In the realm of design, innovative concepts hold paramount importance. With the continuous advancement of image generation technology, designers naturally gravitate towards leveraging artificial intelligence tools like DALL-E~\cite{ramesh2021zero}, Stable Diffusion~\cite{rombach2022high}, and Midjourney~\cite{midjourney} to enhance their creative processes. However, this introduces the potential for data privacy breaches at every stage, from concept and development to utilization. In response to this challenge, we have devised a framework that seamlessly combines a cloud-based large generation model with local fine-tuning models. This integration effectively safeguards sensitive information, such as designers' personalized sketches and final design projects, shielding user privacy and enhancing data security.

%%%%%%%%%%%%%%%%%%%%%%%%%%%%%% OverView %%%%%%%%%%%%%%%%%%%%%%%%%%%%%%
\section{User Study and System Overview}
\label{sec:overview}

\subsection{User Study}
\label{sec:study}

\subsubsection{Questionnaires}
\begin{figure}[ht]
\centering
\includegraphics[width=0.8\linewidth]{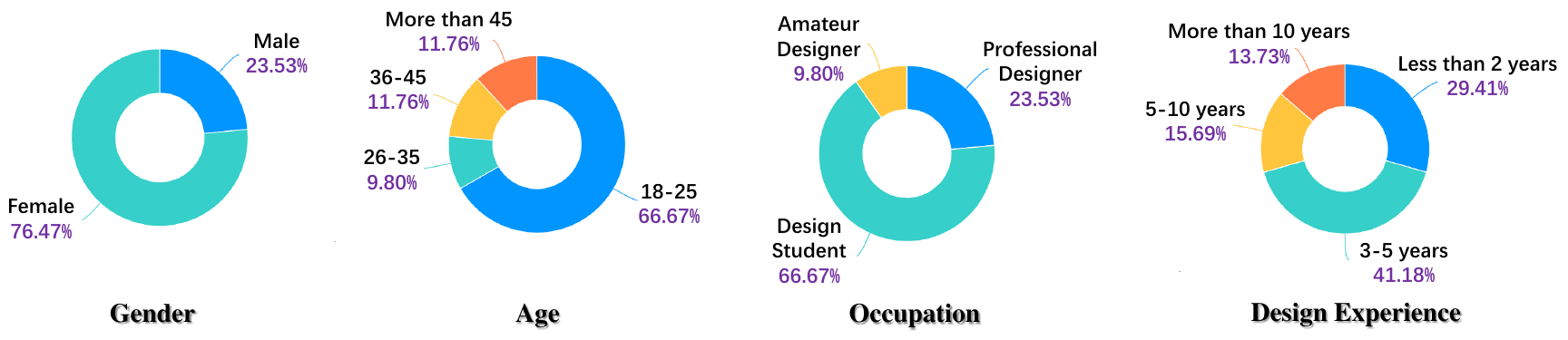}
\caption{Basic information about the respondents.}
\label{fig:ques}
\end{figure}

We invited and collected responses from 51 participants by distributing online questionnaires to individuals in the fashion design industry, including design students, design teachers, and professional designers. The questionnaire comprised a total of 17 questions, covering the respondent's basic information, opinions on traditional design tools, and perspectives on AI-assisted design tools, etc. Initially, our focus was on understanding user demographics, encompassing gender, age, occupation, and design experience. As shown in Fig.~\ref{fig:ques}, most of the respondents were female, constituting 76.47\%, with the highest representation in the 18-25 age bracket at 66.67\%. In terms of occupation, professional designers account for 23.53\%. Notably, design students emerged as the predominant group, constituting the majority and representing the prospective driving force in the future fashion design landscape.

\begin{figure}[ht]
\centering
\includegraphics[width=0.86\linewidth]{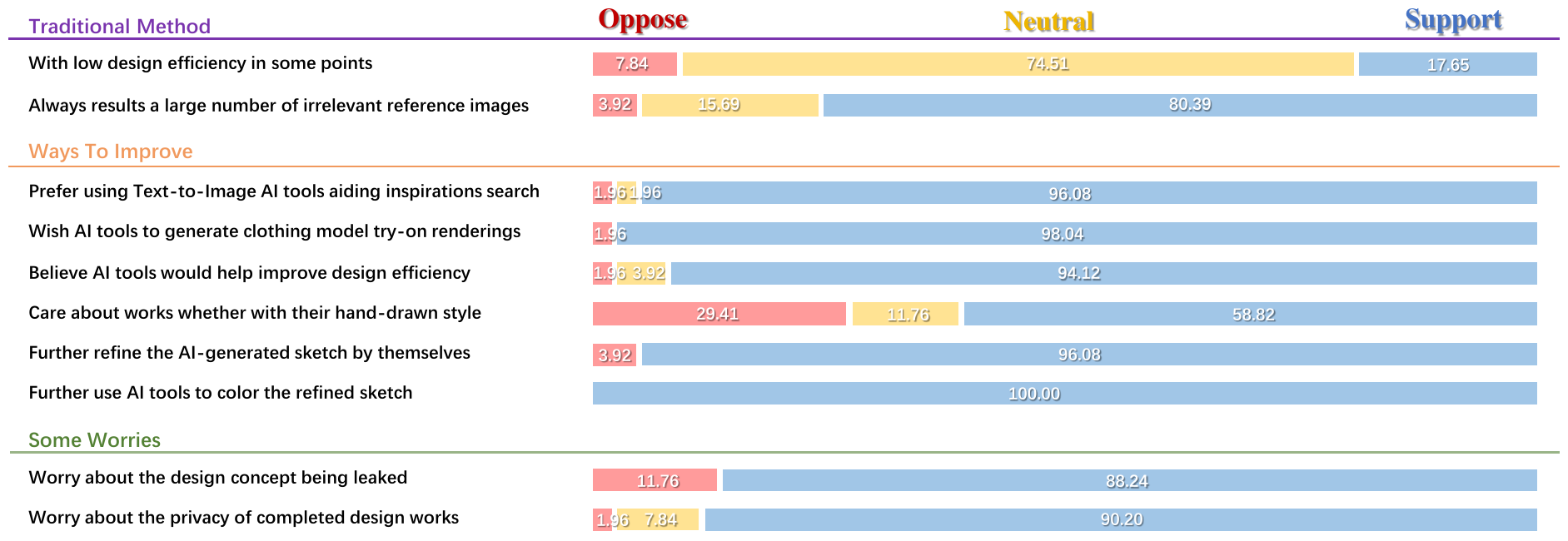}
\caption{Gain insights into the current landscape of fashion designers and unearth their specific needs for AI tools.}
\label{fig:question}
\end{figure}

Subsequent inquiries delved into participants' design practices and perceptions of current design efficiency. We present the main results in Fig.~\ref{fig:question}, a significant 80.39\% indicated reliance on internet searches for design inspiration. And 47.06\% and 41.18\% favored collecting clothing renderings and clothing model try-on images. An overwhelming 80.39\% of respondents indicated that traditional online search methods often yield irrelevant images, this may lead to designers spending significant time on ineffective inspiration retrieval and consequently affecting their overall design efficiency. Further questions probed respondents' perspectives on AI-generated content. An overwhelming 96\% expressed openness to employing an auxiliary design system generating images based on textual input, while 98\% deemed this approach beneficial for obtaining design inspiration. Regarding sketching, 94.12\% believed that automated sketch generation from images would enhance design efficiency, but 58.82\% expressed concerns about the generated sketches lacking their distinct hand-drawn style. Significantly, all respondents expressed a desire to use AI tools for subsequent clothing rendering after generating sketches. Privacy concerns were also prevalent, with 88.24\% worried about the potential leakage of design ideas and 90.2\% expressing concerns about the privacy of their design works. These findings illuminate user perspectives, emphasizing a readiness to embrace AI-driven design tools while underscoring privacy apprehensions.

\subsubsection{Requirements Analysis}
Based on the results of the above questionnaire responses from fashion designers, we have classified and summarized their needs as follows:

\begin{itemize}
\item \textbf{Inspiration Resource.} Generate a plethora of suitable inspiration images for designers using a text-to-image model. Input text can range from brief to detailed descriptions, producing images like clothing renderings and clothing model try-ons. Text prompts allow adjustments to style, background, and other attributes. A novel large model training strategy facilitates rapid learning of new styles.
\item \textbf{Personalized Style.} Utilize a vast collection of fashion images from the internet as a foundational image material library. Designers upload hand-drawn sketches from their personal archives, transforming basic image materials into numerous sketch templates infused with their unique design styles, available for further refinement at any time. Furthermore, designers can continually update the image library over time.
\item \textbf{Coloring Sketch.} Apply a style fusion algorithm based on reference images to showcase the sketch coloring effect, enabling designers to seamlessly blend the refined sketch with the appearance of the inspiration image. This allows designers to rapidly preview the colored sketch and choose the most suitable color style.
\item \textbf{Privacy Protection.} The design processes related to personalized inspiration for designers are deployed on local small models, while the inspiration reference non-personalized design process is deployed on the cloud large model. This approach enables the use of large models to generate impressive inspirational references, while leveraging smaller models to swiftly develop and safeguard local design concepts from potential leaks.
\end{itemize}

\subsection{System Overview}
\label{sec:sys}

\begin{figure}[tb]
\centering
\includegraphics[width=0.68\linewidth]{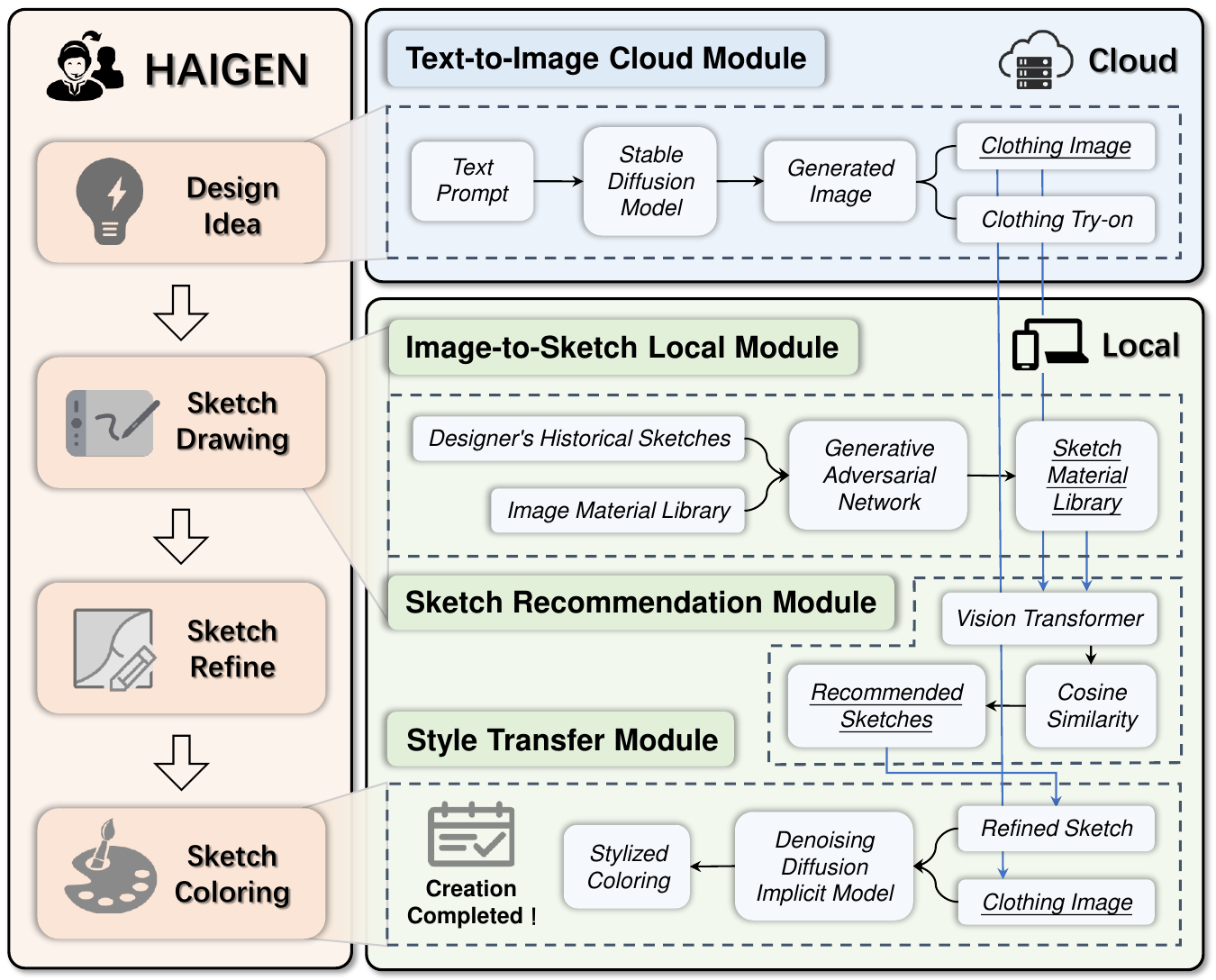}
\caption{An Overview of our HAIGEN system. The left side (orange) illustrates the designer's design process. On the right side, the Text-to-Image Cloud Module (blue) is deployed in the cloud, while the Image-to-Sketch Local Module, Sketch Recommendation Module, and Style Transfer Module (green) are deployed locally.}
\vspace{-2mm}
\label{fig:system}
\end{figure}

The system framework of our HAIGEN is depicted in Fig.~\ref{fig:system}. On the left side (orange) is the designer's design process, which begins with the initial design idea, proceeds to the creation of the first draft, further refinement of the sketch, and ends with the coloring of the sketch. On the right side, the system comprises two main components: the cloud (blue) and the local side (green). The \textit{Text-to-Image Cloud Module}, situated in the cloud, generates images based on detailed sentence descriptions to inspire design. The local side includes the \textit{Image-to-Sketch Local Module}, \textit{Sketch Recommendation Module}, and \textit{Style Transfer Module}, which serve to generate sketch templates, recommend templates to designers, and color refined sketches, respectively. Below, we will provide detailed descriptions of these four modules:

\begin{itemize}
\item \textbf{Stable Diffusion Model-based Text-to-Image Cloud Module.} During the design idea phase, designers often seek inspiration by entering keywords on various online platforms. This process can be time-consuming and arduous, as designers often struggle to find relevant materials. Our method deploys an enhanced Stable Diffusion model~\cite{rombach2022high} in the cloud. Designers can provide detailed descriptions of their ideas, which helps the generated image material closely align with their creative vision. Designers can continuously adjust and refine the generated images, such as altering character backgrounds, appearances, clothing colors, styles, and more. This significantly streamlines the inspiration-finding process,  resulting in the creation of groundbreaking designs that push the boundaries of fashion.

\item \textbf{Capture Personalized Designer-Style Image-to-Sketch Local Module.} In the sketch drawing phase, we employed web crawler technology to establish a foundational image library for designers, accumulating a vast collection of fashion clothing images. Building upon this foundation, we developed an Image-to-Sketch Local Module, utilizing GAN~\cite{goodfellow2014generative} to transform images from the library into sketches reflecting the designer's individual style. This was achieved by learning the overarching style from historical sketch data uploaded by the designer. As a result, designers have access to a comprehensive sketch library that serves as a repository of sketch templates. Furthermore, they have the flexibility to continually expand the library over time.

\item \textbf{Cloud Image-based Local Personalized Sketch Recommendation Module.} When a designer generates a clothing image through the Text-to-Image Cloud Module that they are satisfied with, they can input this image into the Sketch Recommendation Module. This module will recommend similar sketch templates from the library that closely match the clothing image. Designers can then select their preferred sketch template from these recommendations and further refine it to obtain the final sketch. This process significantly simplifies the initial sketching phase, enhancing design efficiency.

\item \textbf{Style Transfer Module based on Local Personalized Sketch and Cloud Image.} During the sketch coloring stage, we've introduced a Style Transfer Module based on DDIM~\cite{song2020denoising}. In this phase, designers input the Refined Sketch and the reference Clothing Image obtained from the previous modules. This module generates images that preserve the unique design style and intricate sketch details of the Refined Sketch while incorporating the color style of the Clothing Image. It's important to note that we ensure privacy protection for designers' sketch ideas by deploying all sketch-related models locally.
\end{itemize}

\subsection{Privacy Protection}
As depicted in Table~\ref{table:privacy}, existing generation methods like Midjourney~\cite{midjourney} and DALL-E~\cite{ramesh2021zero} lack the concept of cloud-local integration. They rely entirely on cloud server resources for computations. Users must upload their data to these servers, where the corresponding content is generated based on the input and then downloaded locally for viewing. This poses a significant privacy risk, as designers are required to upload their private works and inspirations to these servers. DataMix~\cite{liu2020datamix}, while utilizing the cloud-local collaboration framework and leveraging cloud computing resources to alleviate local computing burdens. However, it necessitates the uploading of local data models to the cloud. Despite the introduction of a privacy protection solution, potential privacy concerns persist.

\begin{table}[ht]
\caption{The comparison of several different methods for privacy protection.} 
\label{table:privacy}
\centering
\begin{tabular}{lcccc} \toprule
& Midjourney~\cite{midjourney} & DALL-E~\cite{ramesh2021zero} & DataMix~\cite{liu2020datamix} & HAIGEN (Ours) \\ \midrule
Computation (on Cloud) & 100\% & 100\% & 87\% & 72\% \\
Computation (on Local) & \textcolor[rgb]{0.25, 0.5, 0.75}{0\%} & \textcolor[rgb]{0.25, 0.5, 0.75}{0\%} & 13\% & 28\% \\ \midrule
Download from Cloud & \usym{1F5F8} & \usym{1F5F8} & \usym{1F5F8} & \usym{1F5F8} \\
Upload from Local (Privacy) & \usym{1F5F8} & \usym{1F5F8} & \usym{1F5F8} & \textcolor[rgb]{0.25, 0.5, 0.75}{\usym{2613}} \\ \midrule
% Connection Required & Internet & Internet & Internet & \textcolor{red}{LAN} \\ 
Design Availability & Internet & Internet & Internet & \textcolor[rgb]{0.25, 0.5, 0.75}{LAN} \\ \bottomrule
\end{tabular}
\end{table}

In contrast, our proposed cloud-local collaboration framework strategically allocates tasks based on complexity and privacy considerations. Models requiring extensive computational resources and posing no privacy concerns, such as inspiration image generation models, are deployed in the cloud. Conversely, models related to designer privacy and requiring minimal computational resources are deployed locally. This approach harnesses the potent generation capabilities of large cloud models while eliminating the privacy risks associated with local data. Moreover, concerning Design Availability, outputs generated by techniques like Midjourney~\cite{midjourney} are showcased in a public chat window, enabling others to view the generated results. In contrast, results produced by our approach are exclusively accessible to the respective designer or shared within the Local Area Network (LAN) of the design studio. This discrepancy highlights a crucial distinction in privacy protection between our method and others. Notably, through our framework, designers only need to download inspiration images generated in the cloud, without uploading any image-related private data. Thus, it restricts local data upload, further safeguarding designer privacy.

%%%%%%%%%%%%%%%%%%%%%%%%%%%%%%%%%%%%%%%%%%%%%%%%%%%%%%%%%%%%%%%%%%%%
%%                             Method                             %%
%%%%%%%%%%%%%%%%%%%%%%%%%%%%%%%%%%%%%%%%%%%%%%%%%%%%%%%%%%%%%%%%%%%%
\section{HAIGEN: Towards Human-AI Collaboration for Facilitating Creativity and Style Generation in Fashion Design}
\label{sec:system}
In this section, we will introduce the technical implementation details of our HAIGEN system. It encompasses the \textit{Text-to-Image Cloud Module} (Sec.~\ref{sec:sd}), which visualizes design ideas, the \textit{Image-to-Sketch Local Module} (Sec.~\ref{sec:i2s}), dedicated to generating personalized sketches, the \textit{Sketch Recommendation Module} (Sec.~\ref{sec:sr}), facilitating the recommendation of similar sketch templates, and finally, the \textit{Style Transfer Module} (Sec.~\ref{sec:s2i}), used for sketch coloring.

%%%%%%%%%%%%%%%%%%%%%%%%%%%%%% Stable Diffusion %%%%%%%%%%%%%%%%%%%%%%%%%%%%%%
\subsection{Stable Diffusion Model-based Text-to-Image Cloud Module}
\label{sec:sd}

\begin{figure}[tb]
\centering
\includegraphics[width=0.95\linewidth]{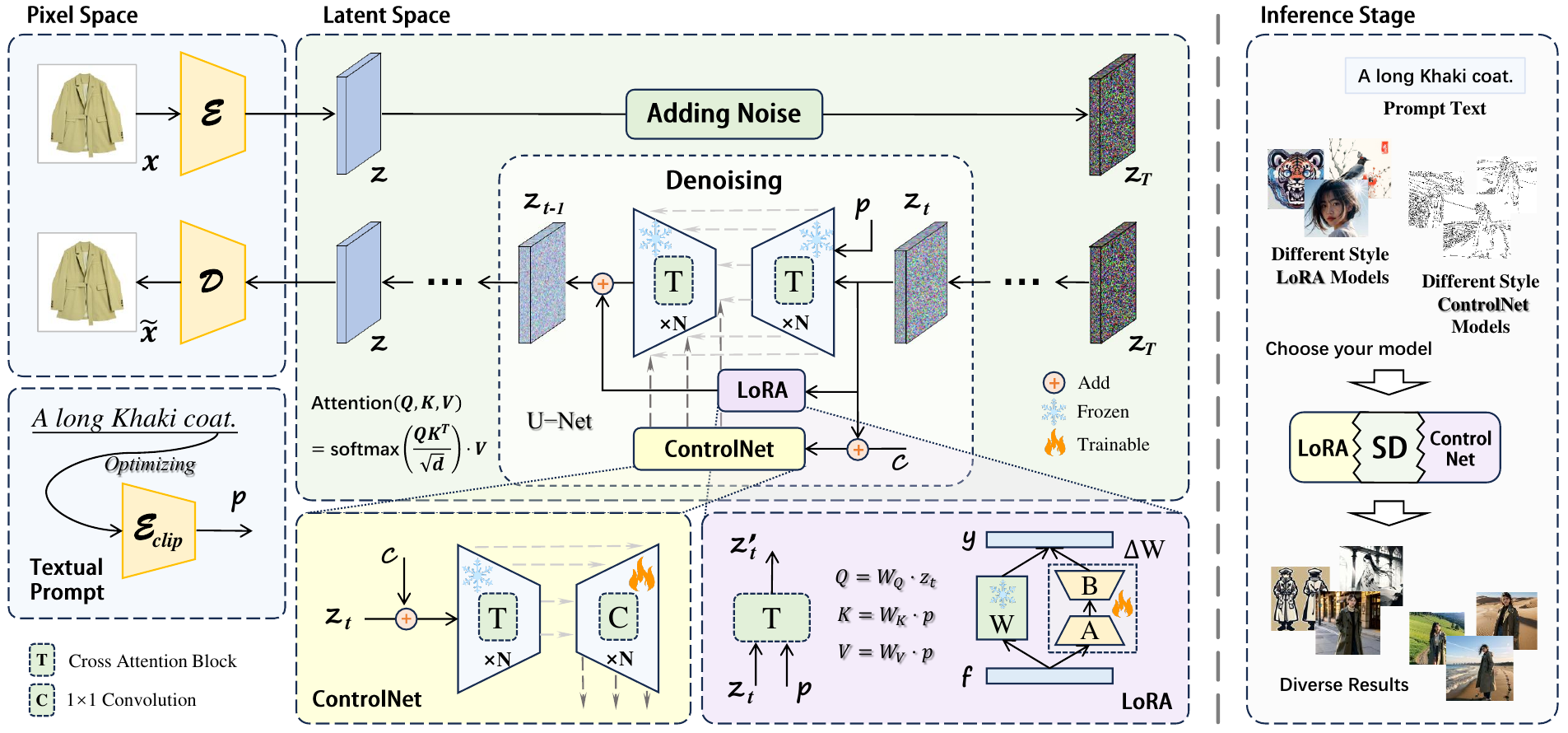}
\caption{The illustration of the Stable Diffusion Model-based Text-to-Image Cloud Module. We initially utilize the VAE~\cite{kingma2013auto} model to map the input sample $x$ into the latent space for diffusion training. To expedite training and exert control over the style and details of the denoising process, we freeze the initial parameters of the SD model and incorporate LoRA~\cite{hu2021lora} and ControlNet~\cite{zhang2023adding}.}
\label{fig:sd}
\end{figure}

In our \textit{Text-to-Image Cloud Module} (T2IM), we generate a variety of high-quality clothing renderings based on text prompts, offering designers an intuitive sense of the designed clothing's quality. As shown in Fig.~\ref{fig:sd}, our T2IM primarily comprises two components: one is a model based on Stable Diffusion (SD) model~\cite{rombach2022high}, and the other involves fine-tuning the SD model using LoRA~\cite{hu2021lora} and ControlNet~\cite{zhang2023adding}. This fine-tuning accelerates the model training process and enhances the detailed information of the generated images.

\subsubsection{Stable Diffusion Model}
Due to the potent image generation capabilities of the Diffusion Model (DM)~\cite{ho2020denoising}, it has exhibited remarkable results in image synthesis and various applications. However, operating in pixel space poses high hardware and time costs for models with extensive parameter quantities. In contrast to DM, the SD model employs a variational autoencoder (VAE)~\cite{kingma2013auto} for encoding and decoding input images based on DM. This addresses the challenge by mapping the input image into a potential low-dimensional representation space for diffusion operations.

In the DM model, image generation is simplified to the process of noise adding and denoising, where all noise follows a Gaussian distribution. Assuming a total of $T$ noise increments are added to the input image $x$, with each noise value represented as $\epsilon_t$ (where $t \in (1, \dots, T)$), the model $\theta$ is trained to predict the noise value $\epsilon_{\theta}(x_t, t)$ that closely matches $\epsilon_t$ at each time $t$. Here, $x_t$ is represented by the result of iteratively adding noise to $x$ over the $t$-th time. We have:
\begin{equation}
\label{eq:DM}
L_{DM} = \mathbb{E}_{x,\epsilon\sim\mathcal{N}(0,1),t}\Big[\|\epsilon-\epsilon_\theta(x_t,t)\|_2^2\Big].
\end{equation}

In the SD model, besides the noise adding and denoising process similar to that in DM, it includes an additional step of encoding and decoding the input image $x$ within the latent space through the VAE model $\mathcal{E}$. And for conditional diffusion with text input, we feed the text prompt into the CLIP~\cite{radford2021clip} text model $\mathcal{E}_{clip}$ to generate feature vector $\boldsymbol{p}$. Throughout the SD model training process, the feature vector is incorporated into the model's diffusion process through the cross-attention module in U-Net, enabling control over the generated image:
\begin{equation}
\label{eq:SD_C}
\mathcal{L}_{SD_T} = \mathbb{E}_{\mathcal{E}(x), \boldsymbol{p}, \epsilon\sim\mathcal{N}(0,1), \boldsymbol{t}} \bigg[\|\epsilon-\epsilon_\theta(\boldsymbol{z}_t,\boldsymbol{t},\boldsymbol{p}))\|_2^2\bigg],
\end{equation}

\noindent where $z_t$ represents the outcome after adding noise to $z$ for the $t$-th iteration.

\subsubsection{SD Model Fine-tuning}
A crucial paradigm in natural language processing involves large-scale pre-training on general domain data, followed by adaptation to specific tasks or domains. However, when dealing with extremely large pre-trained models, complete fine-tuning of all model parameters through retraining becomes infeasible. To address this challenge, Hu \textit{et al.}~\cite{hu2021lora} introduced the low-rank adaptation (LoRA) by preserving the pre-trained model weights and introducing a trainable rank decomposition matrix into each layer of the Transformer~\cite{vaswani2017attention} architecture. This approach significantly reduces the number of trainable parameters for task-specific adaptation. Inspired by LoRA, our fine-tuning of the SD model involves training only the model residuals, utilizing a low-rank matrix to represent the learned content. This not only accelerates model training but also enables diverse transformations of images in specific styles through a combination of various pre-trained LoRA models and SD weights during the inference process, providing a plug-and-play solution. Specifically, we conducted retraining of the weights associated with $Q$, $K$, and $V$ within the cross-attention block of the SD model. As illustrated in Fig.~\ref{fig:sd} [LoRA], during the model training process, we freeze the original weight values $W_0 \in \mathbb{R}^{d \times k}$ while incorporating the concept of residual to introduce a new branch utilizing a low-rank matrix. By utilizing low-rank matrices, we were able to capture new style features and derive a new set of weight values $\Delta W$. Subsequently, during the model inference process, we effectively managed to control the generated image style by integrating the feature outputs from these two sets of weights:
\begin{equation}
\label{eq:lora}
y = W_0f + \Delta Wf = W_0f + BAf,
\end{equation}

\noindent where $A$, $B$ are a set of low-rank matrices, and $A \in \mathbb{R}^{d \times r}$, $B \in \mathbb{R}^{r \times k}$ with the rank $r\ll\min(d,k)$.

In the realm of GAN models, StyleGAN~\cite{karras2020analyzing} stands out. It allows for precise control over specific details of the generated image by manipulating corresponding portions of the input latent vector. Each dimension of the latent vector corresponds to a distinct feature of the generated image, such as facial expression, hairstyle, age, and more. More recently, ControlNet~\cite{zhang2023adding} has presented a similar concept, extending the ability to fine-tune the SD model. It introduces spatially localized input conditions to the pre-trained text-to-image diffusion model, enabling the generation of various image types with detailed feature control. As depicted in Fig.~\ref{fig:sd} [ControlNet], our approach involves the freezing of all weights within the initial SD model. Initially, we duplicate the Encoder from the U-Net architecture~\cite{ronneberger2015u-net} within the SD model and subsequently freeze it. Each layer's output from the Encoder is connected to a 1×1 convolution layer initialized with a weight of 0. We then add the output of each layer to the input of the corresponding layer in the Decoder. Throughout the model training process, we only need to manage a single input perturbation while training the 1×1 convolutional layer. This approach fine-tunes the generated image and allows for precise control over the designated features $c$. The objective function can be succinctly expressed as:
\begin{equation}
\label{eq:control}
\mathcal{L}_{Control} = \mathbb{E}_{\mathcal{E}(x), \boldsymbol{c}, \epsilon\sim\mathcal{N}(0,1), \boldsymbol{t}} \bigg[\|\epsilon-\epsilon_\theta(\boldsymbol{z}_t,\boldsymbol{t},\boldsymbol{c}))\|_2^2\bigg].
\end{equation}

To summarize, during model training, we utilize the LoRA to govern the overall style and the intended object within the generated image. Meanwhile, ControlNet regulates other detail features of the resulting image. Through training on diverse style datasets, we obtain a range of LoRA and ControlNet models. Subsequently, during model inference, we ensure stable control over production images and generate a diverse set of high-quality inspiration images by combining various LoRA and ControlNet models.

\subsubsection{Loss Function} Therefore, the loss function during the overall model training process can be represented as:
\begin{equation}
\label{eq:loss_sd}
\mathcal{L}_{T2IM} = \mathbb{E}_{\mathcal{E}(x), \boldsymbol{p}, \boldsymbol{c}, \epsilon\sim\mathcal{N}(0,1), \boldsymbol{t}} \bigg[\|\epsilon-\epsilon_\theta(\boldsymbol{z}_t,\boldsymbol{t}, \boldsymbol{p},\boldsymbol{c}))\|_2^2\bigg].
\end{equation}

%%%%%%%%%%%%%%%%%%%%%%%%%%%%%% Image-to-Sketch Generation %%%%%%%%%%%%%%%%%%%%%%%%%%%%%%

\begin{figure}[tb]
\centering
\includegraphics[width=\linewidth]{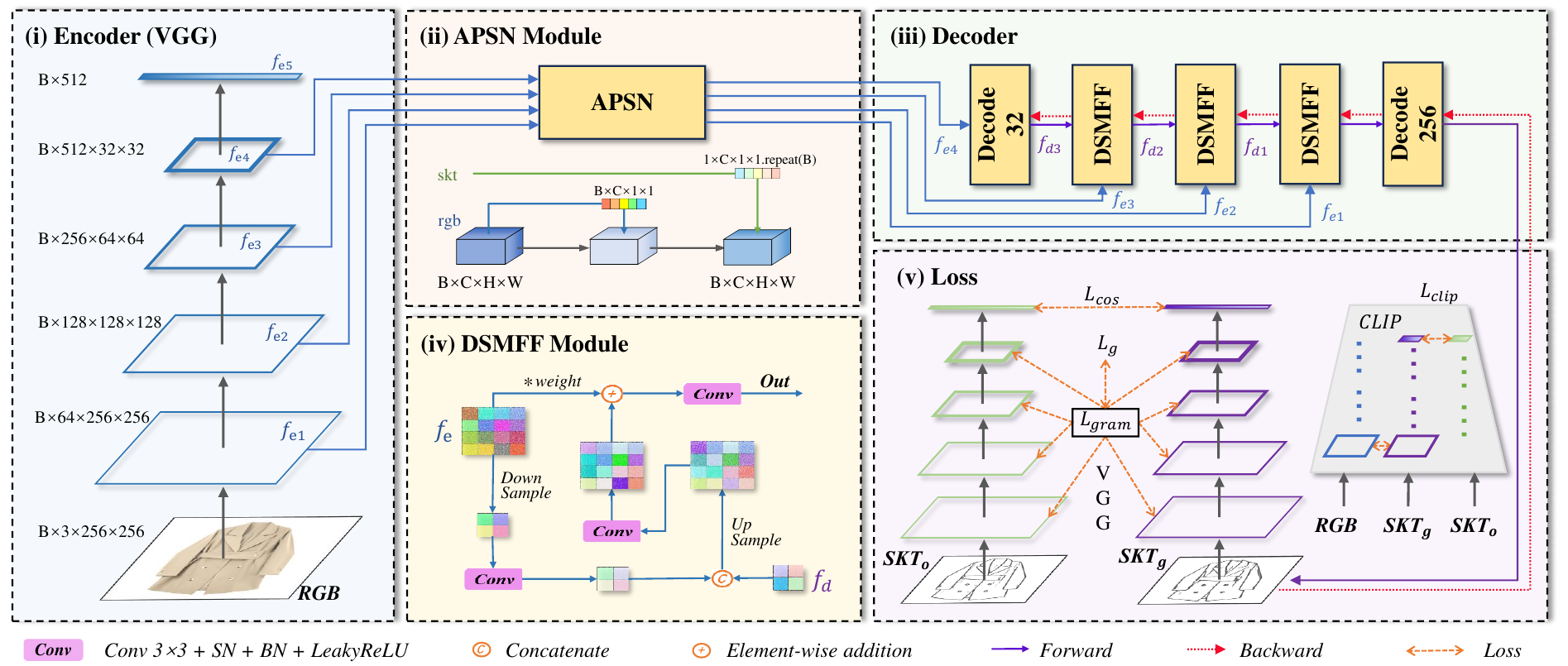}
\caption{The illustration of the Capture Personalized Designer-Style Image-to-Sketch Local Module. We begin by inputting the sketch-image pair ($RGB$ and $SKT_o$) into the system, and multi-level features are extracted using the standard pre-trained VGG-16~\cite{simonyan2014vgg} network. Subsequently, the APSN module and DSMFF module are employed to capture personalized features at multiple levels from the designer's sketch and perform feature fusion to generate the sketch ($SKT_g$) with the designer's style.}
\label{fig:i2s}
\end{figure}

In previous works~\cite{canny1986computational, kim2020u-gat-it, liu2021self}, the task of image-to-sketch generation can be summarized as extracting contour and detail features from the input image. However, our proposed sketch generation model aims to assist designers in quickly completing hand-drawn sketch works, ensuring that the generated sketches possess the unique artistic style of the designer. This is because each designer has their own unique artistic style and conveys their perception of important elements in the sketches from various perspectives through their personal drawing techniques. To illustrate this, we present the framework of our \textit{Image-to-Sketch Local Module} (I2SM) in Fig.~\ref{fig:i2s}, which primarily consists of three components. These components include the standard pre-trained VGG-16~\cite{simonyan2014vgg} for extracting multi-level features, the \textit{Adaptive Personalized Style Normalization} (APSN) Module for capturing and blending different designer styles, and the \textit{DownSampling Multi-Feature Fusion} (DSMFF) Module in Decoder for multi-level feature reconstruction.

\subsection{Capture Personalized Designer-Style Image-to-Sketch Local Module}
\label{sec:i2s}

\subsubsection{VGG Encoder}
The VGG network possesses the advantage of providing rich multi-level feature representation and enhanced feature expression capabilities. It excels in capturing low-level image features, such as edges and textures, in shallower layers, while deeper layers are capable of capturing higher-level semantic features, such as overall shape and structure. We leverage the multi-level feature extraction capability of the VGG-16 to acquire the first four levels of features as input for subsequent sketch reconstruction, namely $f_{e1}$, $f_{e2}$, $f_{e3}$ and $f_{e4}$. The final output linear layer features $f_{e5}$ are utilized to preserve the semantic consistency of the generated output. It should be noted that we select features before the max-pooling layer within the selected first four levels, as the max-pooling operation may lead to the loss of certain local detail features. This ensures that most designer design needs are met as much as possible.

\subsubsection{Adaptive Personalized Style Normalization Module}
The fundamental concept behind the traditional Adaptive Instance Normalization (AdaIN) module is to dynamically adjust the normalization parameters, enabling the model to adapt to distribution variances among different samples and layers. This adaptation enhances the model's generalization capability and adaptability. Building upon this idea, we introduce the \textit{Adaptive Personalized Style Normalization} (APSN) module. It normalizes both the input image and sketch features, followed by mapping the image features utilized for sketch generation to the input sketch feature space. We believe that each designer possesses their own distinctive design style, and as a result, hand-drawn sketches created by the same designer should exhibit similar feature means and variances. This is attributed to the fact that the mean and variance of an image can effectively represent its fundamental characteristics, such as brightness and contrast. Similarly, the mean and variance of sketch features extracted by the VGG encoder can aptly describe the similarity of features, particularly in terms of style-related features.

To be more specific, our objective is to enhance the model's capacity to capture the designer's unique design style by aligning generated sketch features with the characteristics of hand-drawn sketch features. Thus, before model training, we compute the mean $\mu(f_e)$ and variance $\sigma(f_e)$ across all sketch samples from a specific designer. We calculate these statistics for the 0th, 2nd, and 3rd dimensions of the data, corresponding to batch size, height, and width, respectively. During training, for each minibatch of input image samples, we extract their multi-level features using the VGG encoder. These obtained image features are then normalized using the APSN module. Subsequently, the normalized features are fed into the Decoder network, which is responsible for sketch generation, and the normalized features are as follows:
\begin{equation}
\label{eq:apsn}
\begin{split}
APSN(x, \mu_s, \sigma_s) = \sigma_s(\frac{x - \mu(x)}{\sigma(x) + eps}) + \mu_s , \quad f^i_{e_n} = APSN(f^i_{e_n}, \mu(f^s_{e_n}), \sigma(f^s_{e_n})),
\end{split}
\end{equation}

\noindent where $n$ = 1,2,3,4, $eps = {10}^{-6}$ for numerical stability to prevent $\sigma(x)=0$, $f^i_{e_n}$ represents the multi-level features of the input image samples, $f^s_{e_n}$ represents the multi-level features of all sketch samples. It is worth noting that the normalization operation (denoted as "$\mu(x)$" and "$\sigma(x)$" here) is specifically applied to the 2nd and 3rd dimensions of the input feature, which correspond to the height and width dimensions. The purpose of this normalization is to normalize each individual image feature within each minibatch, rather than calculating the mean and variance of all samples collectively.

\subsubsection{DownSampling Multi-Feature Fusion Module}
In the decoder part, we aim to utilize the previously acquired normalized multi-level features for feature decoupling, ensuring that fine-grained information is retained as much as possible. This approach facilitates the generation of high-quality sketches in the designer's style. To achieve this, we introduce the \textit{DownSample Multi-Feature Fusion} (DSMFF) module, illustrated as part (iii) and (iv) in Fig.~\ref{fig:i2s}.

For the input features $f_e$ and $f_d$, their respective dimensions are determined as $f_e \in \mathbb{R}^{B \times C \times W \times W}$ and $f_d \in \mathbb{R}^{B \times C \times W/2 \times W/2}$. Here, $f_e$ represents the feature normalized by the APSN module, $f_d$ represents the feature decoded by the previous level in the Decoder, $B$ represents the batch size of the feature, $C$ represents the number of channels, $W$ represents the size of the feature map. In the beginning, we downsample (DS) the input feature $f_e$ by using max-pooling. The DS allows us to obtain a feature map that captures more global semantic information, thereby capturing the overall structure of the input feature. Following this, we utilize a convolution layer, which is composed of a 3$\times$3 convolution kernel. Additionally, we apply 1-pixel padding convolution to maintain the spatial dimensions of the feature map. To enhance the stability and performance of the network, we incorporate spectral normalization (SN)~\cite{miyato2018spectral} and batch normalization (BN)~\cite{ioffe2015batch}. And employ the LeakyReLU activation function to introduce non-linearity and avoid vanishing gradient, so the convolution layer $Conv(x) =  LeakyReLU(BN(SN(Conv_{3\times3} (x))))$. Next, we concatenate this downscaled feature with the decoded feature $f_d$. Subsequently, we perform upsampling (US) and convolution operations $Conv(x)$ to generate a feature that incorporates both the encoding information from the image layer and the decoding information from the upper layer. Finally, we add this combined feature to the input feature $f_e$. To ensure that the resulting feature contains both global information from the combined feature and fine-grained information from $f_e$. Furthermore, we introduce a differentiable parameter $\gamma$ as the weight of the $f_e$ feature to minimally affect the final decoded features. In summary, the described process can be mathematically expressed as:
\begin{equation}
\label{eq:decode}
\begin{split}
D1(f_e) &= Conv(DS(f_e)), \quad
D2(f_e, f_d) = Conv(US(Cat(D1(f_e), f_d))) \\
DSMFF(f_e, f_d) &= Conv(D2(f_e, f_d) + \gamma \cdot f_e), \quad
f_{d_{n-1}} = DSMFF(f_{e_n}, f_{d_n}), n=3,2,1.
\end{split}
\end{equation}

\subsubsection{Loss Function}
Our loss function comprises four components: $\mathcal{L}_{gram}$, $\mathcal{L}_{g}$, $\mathcal{L}_{cos}$, and ${L}_{clip}$. Among them, $\mathcal{L}_{gram}$ uses the Gram matrix and Mean Square Error loss to control the consistency of detailed features in the generated sketches. $\mathcal{L}_{g}$ is the discriminator loss, evaluating the quality of the generated sketches and facilitating model convergence. $\mathcal{L}_{cos}$ and CLIP-based~\cite{radford2021clip} ${L}_{clip}$, control the semantic feature consistency and global feature consistency of the reference sketch and the generated sketch, respectively. The weight values assigned to these losses are as follows: $\lambda_{gram}$=100.0, $\lambda_{g}$=1.0, $\lambda_{cos}$=0.5, $\lambda_{clip}$=100.0. These weight values are set to achieve a balanced and stable training process, ensuring that each loss function contributes effectively to the overall model optimization. Therefore, the complete loss function of our \textit{Image-to-Sketch Generation} model is defined as:
\begin{equation}
\label{eq:lossis}
\mathcal{L}_{I2SM} = \lambda_{gram}\mathcal{L}_{gram}+\lambda_{g}\mathcal{L}_{g}+\lambda_{cos}\mathcal{L}_{cos}+\lambda_{clip}\mathcal{L}_{clip}.
\end{equation}

%%%%%%%%%%%%%%%%%%%%%%%%%%%%%% Sketch Recommendation %%%%%%%%%%%%%%%%%%%%%%%%%%%%%%
\subsection{Cloud Image-based Local Personalized Sketch Recommendation Module}
\label{sec:sr}

\begin{figure}[tb]
\centering
\includegraphics[width=0.98\linewidth]{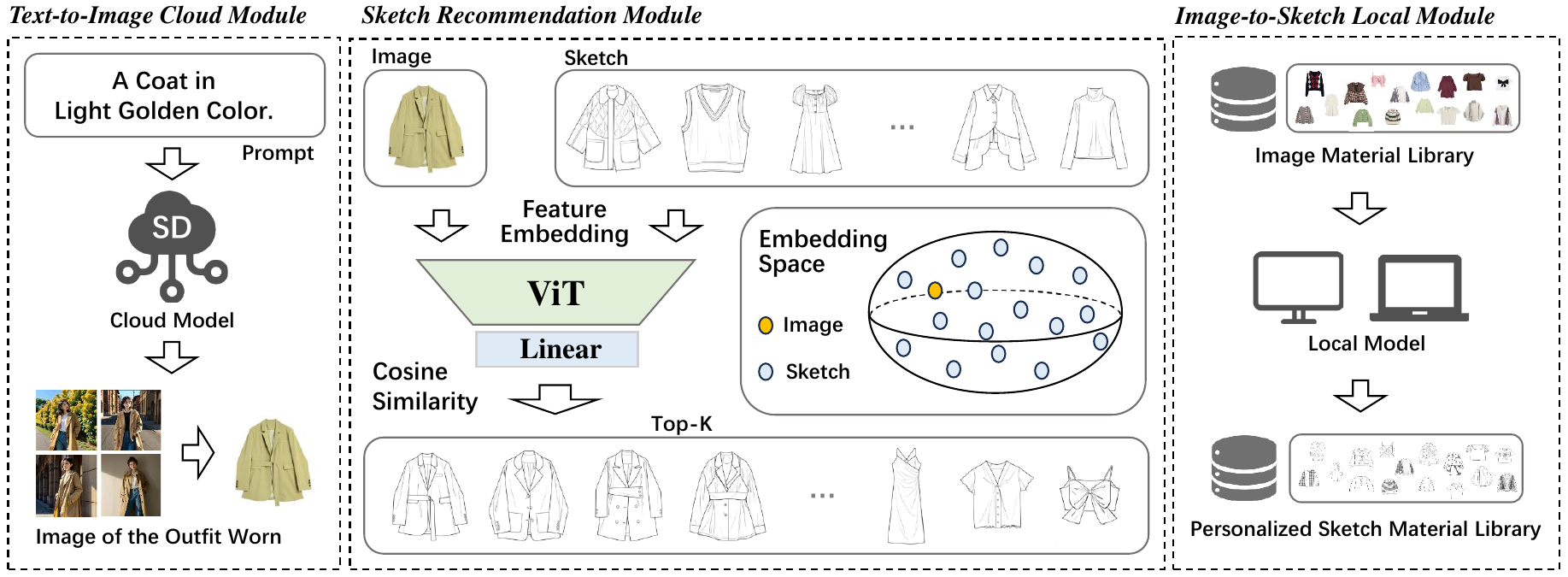}
\caption{The illustration of the Cloud Image-based Local Personalized Sketch Recommendation Module. We first acquire the inspiration image and personalized sketch material library through the \textit{Text-to-Image Cloud Module} and \textit{Image-to-Sketch Local Module}. Subsequently, we leverage the ViT~\cite{dosovitskiy2020vit} to assess their similarity and output recommendations based on the resulting rankings.}
\label{fig:sr}
\end{figure}

In this module, our goal is to identify sketches within the local sketch material library that closely match the input cloud image, thereby further improving the existing materials and finally realizing sketch-based image style transfer for coloring. Illustrated in Fig.~\ref{fig:sr}, we present the workflow of the personalized sketch recommendation system. It involves obtaining a substantial collection of personalized sketch material libraries through a locally pre-trained sketch-to-image generation model and image material library (Sec.~\ref{sec:i2s}). We generate numerous images of the outfit worn in various scenes through a cloud-based pre-trained text-to-image generation model and the input reference text prompt (Sec.~\ref{sec:sd}). Then, we retrieve the corresponding clothing image and recommend the most similar sketch materials from the personalized sketch material library based on the clothing image. Designers can further refine the recommended sketch materials and input them into our Style Transfer Module (Sec.~\ref{sec:s2i}).

\subsubsection{Sketch Recommendation Module}
The task of image-based sketch recommendation can be categorized as cross-modal data retrieval since sketches and images represent the same objects but exhibit distinct modal characteristics. Numerous studies~\cite{dosovitskiy2020vit, han2021tnt, liu2021Swin} have demonstrated the superior feature extraction capabilities of Vision Transformer (ViT) models, especially concerning global feature extraction from image data when compared to Convolutional Neural Networks (CNN) models. In the context of the image-based sketch recommendation task, there is a need to distinguish between various clothing styles, such as shirts, coats, skirts, etc., followed by capturing fine-grained details within the sketches. For this reason, we opted for the standard ViT-B/16 model~\cite{dosovitskiy2020vit} as the feature extraction model for our sketch recommendation module. The workflow begins with the acquisition of extensive personalized sketch materials and reference clothing data, which are obtained through the Image-to-Sketch Generation Module and Text-to-Image Cloud Module. These data are then input into the pre-trained Vision Transformer (ViT) model, followed by a connected linear layer for feature embedding of both sketches and the reference image. The central part of Fig.~\ref{fig:sr} illustrates the distribution of embedded sketches and the reference image within the embedding space. Based on the relative distances in this embedding space, we recommend a Top-K ranking of sketches within the personalized sketch material library. Designers can then use this ranking to select suitable materials for further refinements based on the recommended sketches. Specifically, we adopt cosine similarity to measure the similarity between the input image and the sketch template, which is expressed mathematically as follows:
\begin{equation}
\label{eq:l_clip}
\text{Cosine Similarity}(A, B) = \frac{A \cdot B}{\|A\| \cdot \|B\|},
\end{equation}
\noindent where $A$ and $B$ represent the feature vectors of the sketches and the reference image, $\cdot$ represents the inner product between the vectors, and $\|A\|$ and $\|B\|$ represent their Euclidean norms.

%%%%%%%%%%%%%%%%%%%%%%%%%%%%%% Sketch-to-Image Style Transfer %%%%%%%%%%%%%%%%%%%%%%%%%%%%%%
\subsection{Style Transfer Module based on Local Personalized Sketch and Cloud Image}
\label{sec:s2i}

\begin{figure}[tb]
\centering
\includegraphics[width=0.95\linewidth]{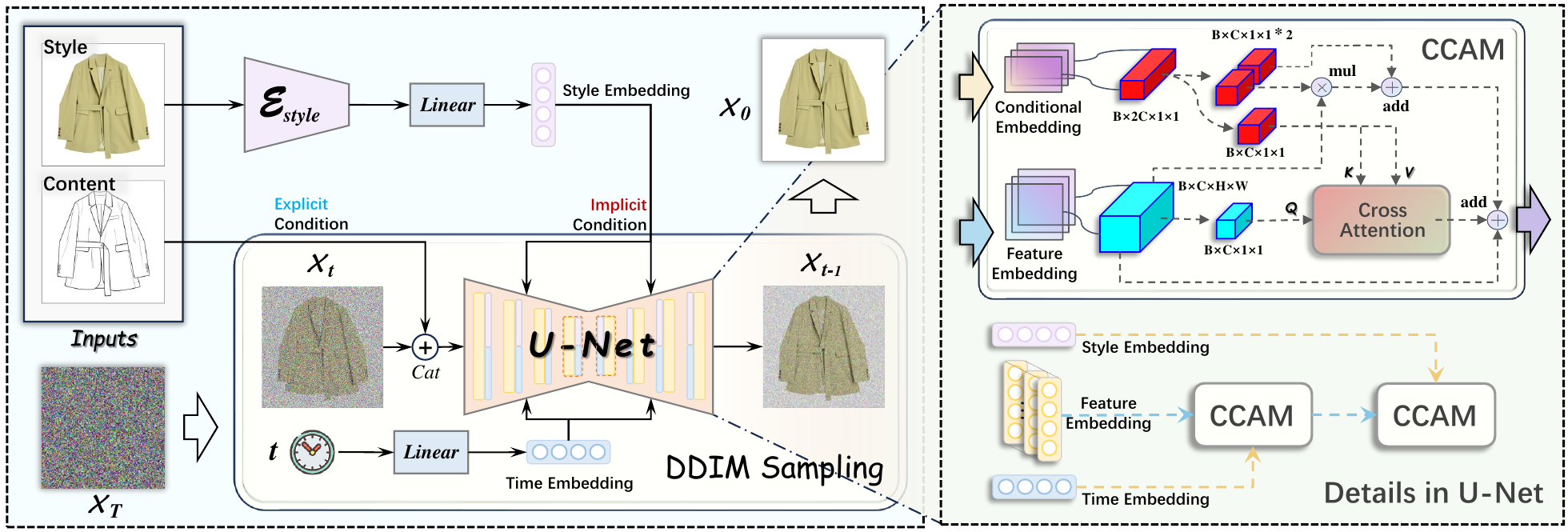}
\caption{The illustration of the Style Transfer Module based on Local Personalized Sketch and Cloud Image. We incorporate the sketch and reference image into the DDIM~\cite{song2020denoising} diffusion process through both explicit and implicit ways to supervise model generation. This integration, coupled with our proposed CCAM, enhances the overall quality of the generated image.}
\label{fig:s2i}
\end{figure}

At this stage, our proposed Style Transfer Module (STM) merges the sketch obtained locally refined with the image acquired from the cloud model, generating a customized sketch image imbued with the stylistic elements of the cloud image, as present in Fig.~\ref{fig:s2i}. This approach empowers designers to quickly apply personalized color to their sketches, facilitating the selection of the most fitting color style for their work and thereby enhancing designer efficiency. The use of a private, locally located sketch ensures that the designer's creative ideas remain secure, effectively safeguarding their privacy. Our STM is built around the core component of the Denoising Diffusion Implicit Model (DDIM)~\cite{song2020denoising}, with the generation of DDIM guided by the addition of sketches and images as content and style conditions. Furthermore, we enhance the conditions for the diffusion process through our proposed \textit{Channel Cross Attention Module} (CCAM), ensuring the production of high-quality, realistic images.

\subsubsection{Denoising Diffusion Implicit Model}
Differing from traditional Generative Adversarial Networks (GANs)~\cite{goodfellow2014generative}, the Denoising Diffusion Probability Model (DDPM)~\cite{ho2020denoising} transforms the image generation problem as a process of noise adding and denoising. It generates realistic images by continually denoising pairs of random Gaussian noise. In the forward noise-adding process, at time $t$ $(1\leq t \leq T)$, the image $x_t$ is defined as $x_t = \alpha_t x_0 + \sqrt{1 - \alpha_t^2} \epsilon_t$, where $\alpha_t$ is a gradually decreasing hyperparameter ensuring that $x_T$ approximately equal to Gaussian distribution as $T$ becomes sufficiently large. During the model inference process, noise added during the forward process is predicted and gradually removed to obtain the generated image. In our method, we adopt DDIM~\cite{song2020denoising} as our sampling method because it overcomes the longer generation times issue associated with the Markov chain in the DDPM reverse denoising process.

Taking inspiration from DiffusionRig~\cite{ding2023diffusionrig}, we introduce sketches and images as conditional supervision to guide the model's generation process. This conditional supervision is divided into explicit and implicit conditions. For explicit conditions, we directly concatenate the sketch $\boldsymbol{c}$ and noise image $x_t$ as the input to the U-Net~\cite{ronneberger2015u} denoising model at time $t$. This ensures that the final generated image seamlessly fits the input sketch outline and preserves as much sketch detail as possible. For implicit conditions, we encode the style of the reference image $\boldsymbol{s}$ and the time $t$ of the current diffusion step into the model's denoising process using the cross-attention mechanism inspired by CrossViT~\cite{chen2021crossvit}. This ensures that the generated image possesses the same style characteristics as the reference image, and leverages time encoding to enhance the model's ability to predict noise during each denoising pass. For encoding the style of the reference image, we use features obtained through ResNet~\cite{he2016deep} with a linear layer, and for encoding the diffusion time, features are acquired through two linear layers. Consequently, the optimization objective of our model can be expressed as:
\begin{equation}
\label{eq:ddim}
\mathcal{L}_{STM} = \mathbb{E}_{\boldsymbol{c}, \boldsymbol{s}, \epsilon\sim\mathcal{N}(0,1), \boldsymbol{t}} \bigg[\|\epsilon-\epsilon_\theta((\boldsymbol{x}_t,\boldsymbol{c}), \boldsymbol{s},\boldsymbol{t}))\|_2^2\bigg],
\end{equation}
\noindent where $\epsilon_\theta$ is the denoising model to predict the noise $\epsilon$ added during the forward diffusion process.

\subsubsection{Channel Cross Attention Module}
In the noise prediction U-Net model $\epsilon_\theta$, we introduce the \textit{Channel Cross Attention Module} (CCAM) to maintain stable control over the denoising process through conditional embedding, as illustrated in the right part of Fig.~\ref{fig:s2i}. Differing from previous methods~\cite{rombach2022high, ding2023diffusionrig}, we sequentially add time embedding and style embedding to the U-Net, with our cross-attention module primarily focusing on channel information within the feature map. Specifically, we split the input conditional embedding and feature embedding into two branches. One achieves rapid feature fusion through matrix multiplication and addition, efficiently retaining more features from both inputs. The other utilizes channel cross attention, obtaining channel information maps through feature dimensionality reduction and concentration. Subsequently, we employ feature fusion of their channel information to facilitate the learning of channel information. This methodology of CCAM is applied to both time embedding and style embedding, enhancing feature fusion capabilities, and eventually yielding fused features as output. It is noteworthy that we selectively added the CCAM to certain layers to improve model training and inference efficiency, while ensuring optimal performance in model generation.

%%%%%%%%%%%%%%%%%%%%%%%%%%%%%%%%%% Experiments %%%%%%%%%%%%%%%%%%%%%%%%%%%%%%%%%%
\section{Experiments}
\label{sec:experiments}
In this section, we will discuss the experimental results and ablation studies to demonstrate the contribution of different components. Additionally, we will compare our proposed system, HAIGEN, with different state-of-the-art approaches through quantitative and qualitative evaluations.

\subsection{Implementation Details}
We implemented our model using PyTorch, and our experiments were conducted on the open-source clothing dataset HAIFashion~\cite{jiang2024haifit}, which comprises 3,100 fashion clothing images. For the \textit{Text-to-Image Cloud Module} (Section~\ref{sec:sd}) experiments, we utilized the NVIDIA Tesla A100 40G GPUs. The optimizer used was AdamW with a learning rate of 1e-4, a batch size of 8, and the scheduler was set to Constant with Warmup. The model was trained for 100 epochs with 8-bit quantization. Local module experiments were performed on a single NVIDIA Tesla V100 32G GPU. In the \textit{Image-to-Sketch Local Module} (Section~\ref{sec:i2s}), we used the SGD optimizer with an initial learning rate of 2e-4 and a momentum of 0.9. The batch size was set to 8, and each training session used 100 pairs of sketch-image data from three different designers in Clothes-V1~\cite{jiang2024simple}, totaling 50 epochs. For the \textit{Sketch Recommendation Module} (Section~\ref{sec:sr}), we employed the pre-trained ViT-B/16 model~\cite{dosovitskiy2020vit}. In the \textit{Style Transfer Module} (Section~\ref{sec:s2i}), we used the Adam optimizer with an initial learning rate of 1e-4 and a batch size set to 8, and the model was trained for 500,000 iterations. The dataset is partitioned into 2,500 pairs for the train and 600 pairs for the test.

\subsection{Baselines}
We conducted comparative experiments using our proposed method and other baselines. We Use the same preprocessing steps and parameters provided by the original authors for each model to ensure a fair and consistent comparison.
\textbf{Canny}~\cite{canny1986computational} generates edge maps of images through thresholding and edge pixel concatenation.
\textbf{pix2pix}~\cite{isola2017image} pioneered a comprehensive solution for image translation using conditional generative adversarial networks.
\textbf{UGATIT}~\cite{kim2020u-gat-it} combines a novel attention module and a learnable normalization function.
\textbf{Self-Sup}~\cite{liu2021self} introduces an unsupervised image-to-sketch generation method and proposes a multi-stage style transfer method.
\textbf{StyleMe}~\cite{wu2023styleme} guides the edge feature generation of the image through class activation mapping and ensures the consistency of style transfer by separating the styles in the generated images.
\textbf{AdaIN}~\cite{huang2017adain} achieves the style transfer by applying instance normalization to each channel pixel.
\textbf{UCAST}~\cite{zhang2023unified} learns style representations directly from massive images through contrastive learning.
\textbf{DiffusionRig}~\cite{ding2023diffusionrig} proposes an image editing method based on the diffusion model and 3D prior.

\subsection{Evaluation Metrics}
We employ five commonly used quantitative perceptual evaluation metrics to assess the objective quality of generated images:
\textbf{PSNR} (Peak Signal-to-Noise Ratio) measures the relationship between signal and noise in the reconstructed and original images based on the mean square error.
\textbf{SSIM} (Structural Similarity Index)~\cite{wang2004image} evaluates the similarity of the original image and the reconstructed image by considering their brightness, contrast, and structural information.
\textbf{MSE} (Mean Squared Error) compares the differences between images by calculating the average of the squared differences of each pixel between the original image and the reconstructed image.
\textbf{LPIPS} (Learned Perceptual Image Patch Similarity)~\cite{zhang2018unreasonable} utilizes deep convolutional neural networks to learn perceptual differences, providing a metric that better simulates human visual perception.
\textbf{FID} (Fréchet Inception Distance)~\cite{heusel2017gans} measures the difference in feature distribution and statistics between generated samples and real samples, offering a reliable measure consistent with human subjective perception.

%%%%%%%%%%%%%%% Text-to-Image Cloud Module %%%%%%%%%%%%%%%%%%%
\subsection{Performance of Text-to-Image Cloud Module}
\subsubsection{Optimization Results}

\begin{figure}[ht]
\begin{center}
\includegraphics[width=0.9\linewidth]{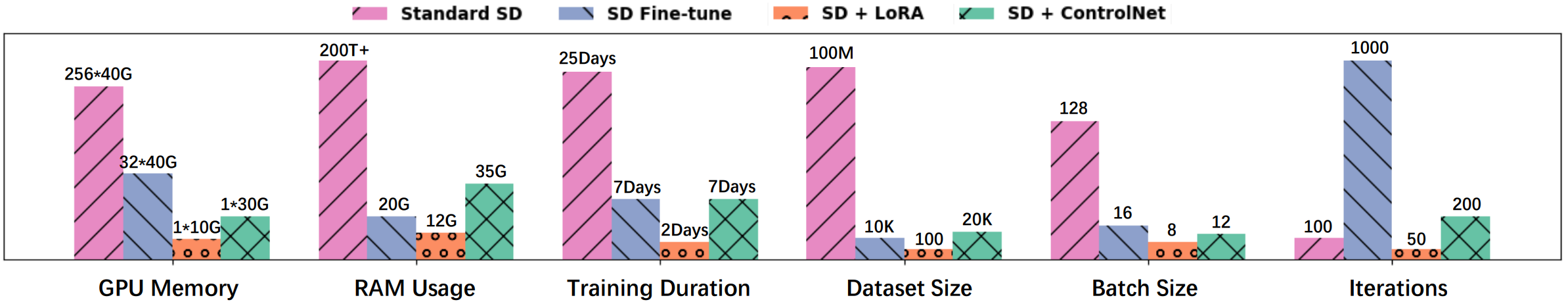}
\caption {The comparison of various demands brought by different SD training methods.}
\label{fig:sd_com}
\end{center}
\vspace{-3mm}
\end{figure}

We compare the performance of four approaches, namely Stable Diffusion (SD) model training, SD Fine-tune, SD+LoRA, and SD+ControlNet, from six perspectives: GPU memory consumption, RAM usage, training duration, dataset size, batch size, and training iterations during the training process, which as shown in Fig.~\ref{fig:sd_com}. Our findings reveal that the traditional methods of direct SD training or utilizing the SD model for fine-tuning significantly burden computational resources, creating an unfavorable scenario for designers that may hinder their design process. In contrast, our approach, incorporating LoRA and ControlNet, minimizes data, hardware, and time requirements for SD model learning of specific styles. This method not only efficiently satisfies these requirements but also allows designers to swiftly access multiple different LoRA and ControlNet models, offering diverse style effects through random combinations during usage.

\vspace{-3mm}
\begin{figure}[ht]
\begin{center}
\includegraphics[width=0.92\linewidth]{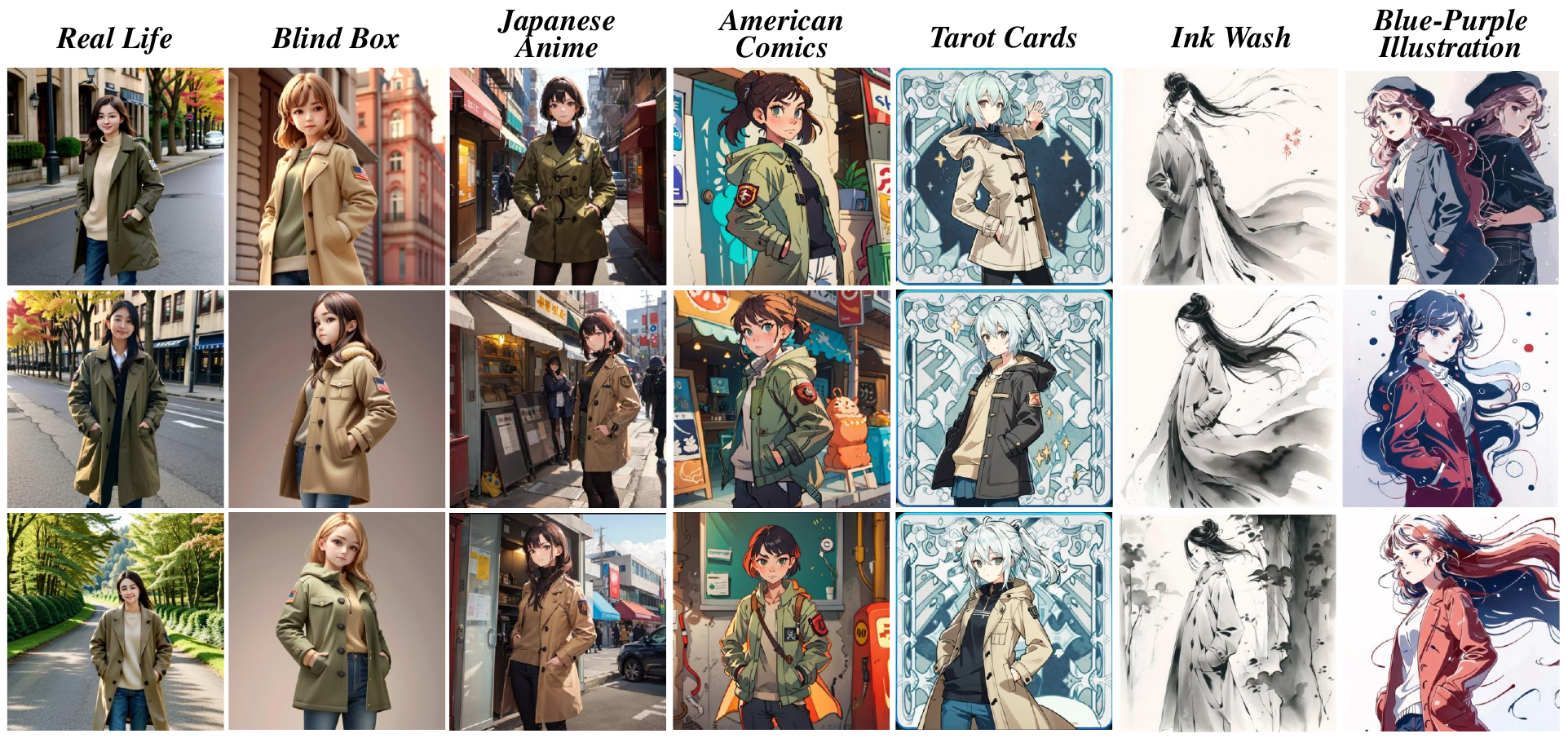}
\caption {Given a consistent text prompt as input, a diverse set of satisfying inspiration images are generated by selecting different styles (each column represents images in the same style).}
\vspace{-3mm}
\label{fig:t2i_gen}
\end{center}
\end{figure}

\subsubsection{Qualitative Results}
Fig.~\ref{fig:t2i_gen} illustrates the performance of our system for generating images from text prompts. Using a consistent input text, "A girl wears a khaki coat and posed with her hands in her pockets," we combine the generated inspiration images with LoRA models featuring different styles. Regardless of the chosen style, the results closely align with the description of "girl," "khaki coat," and "hands in pockets" from the input text. The generated images also exhibit a general consistency with the desired style. The last two columns show a discrepancy because the requested style conflicts with the khaki color characteristics, which hinders the accurate representation of the coat's khaki color. Despite this, other characteristics are well-displayed. Upon closer examination of the generated images, one can observe rich details, including facial features, clothing, and the surrounding environment. These exceptional results are attributed to LoRA's control over the overall image style and ControlNet's further optimization of image details.

%%%%%%%%%%%%%%% Image-to-Sketch Local Module %%%%%%%%%%%%%%%%%%%
\subsection{Performance of Image-to-Sketch Local Module}
\subsubsection{Quantitative Results}
Table~\ref{table:i2s} presents the quantitative analysis of our image-to-sketch generation results, comparing our proposed model with several baseline models and showcasing the results of ablation experiments on relevant modules. We chose the FID index as our measurement standard because it closely aligns with human perception and effectively reflects the quality of sketch generation. HAIGEN demonstrates excellent performance across the three professional designer datasets and outperforms the four baseline models. Moreover, Our model boasts smaller sizes, enabling faster training and inference speeds compared to UGATIT, Self-Sup, and StyleMe. This compact size provides advantages in terms of flexibility, efficiency, and ease of deployment. It sets our model apart from the Stable Diffusion model, making it more practical and scalable for local devices. In the ablation study, we primarily focus on validating the effectiveness of the APSN module and the downsample plus concatenation operation (referred to as "Down" in Table~\ref{table:i2s}) in the DSMFF module. The presence of the APSN module and the Down module significantly enhances the overall performance of our model. This improvement can be attributed to the capability of the APSN module to learn the distinctive style of the input hand-drawn sketches by capturing their overall feature distribution. Additionally, the Down module effectively utilizes the features extracted by the VGG encoder to fuse detailed information from the input features with contour features from the downsampling process. This fusion enables a more refined feature representation, resulting in enhanced output quality with intricate details.

\begin{table}[tb]
\caption{Quantitative experimental results on image-to-sketch generation. ($\downarrow$ indicates the lower value, the better effect.)} 
\label{table:i2s}
\centering
\begin{tabular}{clcccccc} \toprule
\multirow{2}{*}{} & \multirow{2}{*}{\textbf{Methods}} & \multicolumn{1}{c}{\textbf{Model Size}} & \multicolumn{1}{c}{\textbf{Training}} & \multicolumn{1}{c}{\textbf{Inference}} & \multicolumn{3}{c}{\textbf{Clothes-V1} (\textbf{FID} $\downarrow$)} \\
& & (MB) & (hours) & (items/s) & \textbf{Designer1} & \textbf{Designer2} & \textbf{Designer3} \\ \midrule \midrule
\multirow{5}{*}{\rotatebox[origin=c]{90}{Baselines}}
& \multicolumn{1}{l}{Canny~\cite{canny1986computational}} & - & - & - & 245.8723 & 239.1300 & 226.3630 \\
& \multicolumn{1}{l}{pix2pix~\cite{isola2017image}} & 43.52 & 0.76 & 115.93 $\pm$ 0.03 & 135.8482 & \underline{83.0723} & 106.9900 \\
& \multicolumn{1}{l}{UGATIT~\cite{kim2020u-gat-it}} & 1054.72 & 1.08 & 19.46 $\pm$ 0.01 & 154.7540 & 126.8223 & 130.4817 \\
& \multicolumn{1}{l}{Self-Sup~\cite{liu2021self}} & 21.33 & 0.05 & 68.67 $\pm$ 0.02 & 84.8238 & 98.9730 & 97.1567 \\
& \multicolumn{1}{l}{StyleMe~\cite{wu2023styleme}} & 1087.98 & 0.11 & 36.28 $\pm$ 0.05 & \underline{80.5416} & 83.8772 & \underline{88.9189} \\ \midrule
\multirow{4}{*}{\rotatebox[origin=c]{90}{Ablation}}
& \multicolumn{1}{l}{w/o APSN and Down} & 14.34 & \underline{0.04} & 89.28 $\pm$ 0.02 & 95.9745 & 86.6144 & 90.6861 \\
& \multicolumn{1}{l}{w/o APSN} & \underline{13.66} & \underline{0.04} & \textbf{147.79 $\pm$ 0.02} & 98.2104 & 98.7075 & 106.3330 \\
& \multicolumn{1}{l}{w/o Down} & 14.34 & \underline{0.04} & 83.85 $\pm$ 0.01 & 84.4137 & 85.4105 & 89.1427 \\
& \multicolumn{1}{l}{\textbf{HAIGEN (Ours Full)}} & \textbf{13.66} & \textbf{0.04} & \underline{135.25 $\pm$ 0.03} & \textbf{76.9561} & \textbf{78.6799} & \textbf{86.8245} \\ \bottomrule
\end{tabular}
\end{table}

\subsubsection{Qualitative Results}
\begin{figure}[ht]
\centering
\includegraphics[width=\columnwidth]{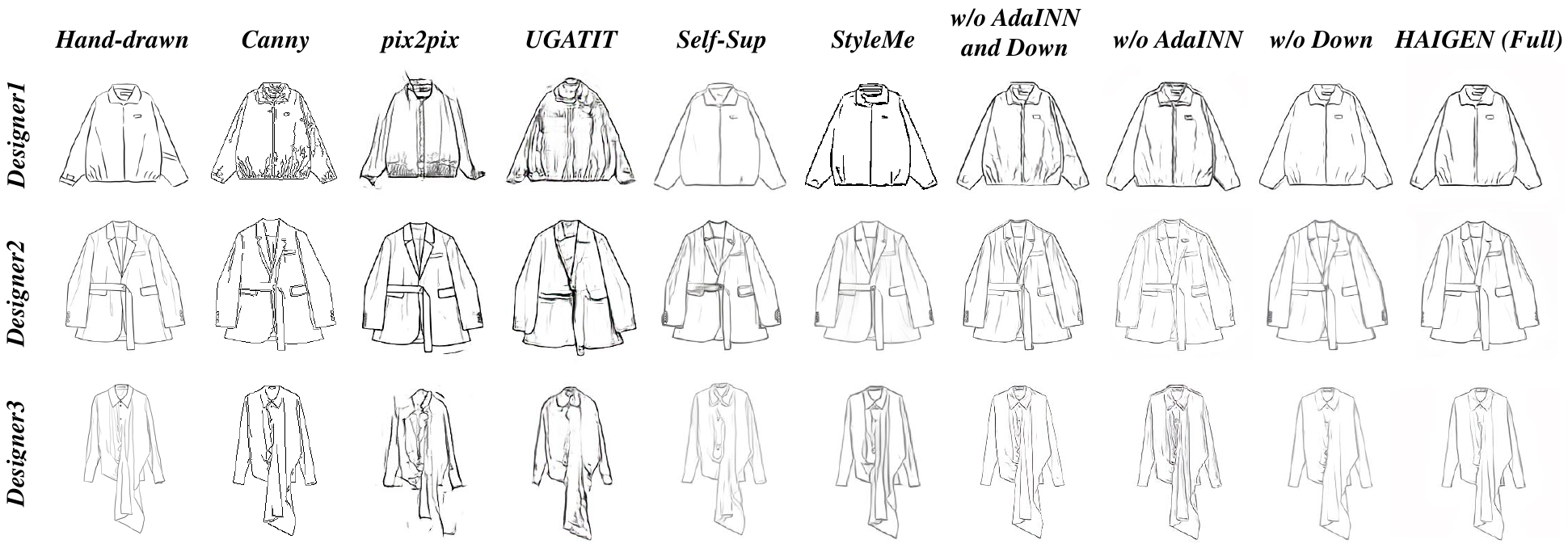}
\caption{Comparison of generated sketches from the manuscripts of three professional designers.}
\vspace{-3mm}
\label{fig:i2s_gen}
\end{figure}

Furthermore, we showcase the qualitative results in the image-to-sketch generation of our proposed model compared with four baselines along with the ablation study of the APSN module and Down module as shown in Fig.~\ref{fig:i2s_gen}. The observed differences in the sketch styles generated by other baseline models indicate certain issues, such as the lines in Canny exhibit severe pixelation, UGATIT's outline edges appear blurred, and while Self-Sup and StyleMe perform better, they still face challenges with excessive details (Designer2) or pixelated lines (Designer1). Similarly, in the model ablation experiments, the absence of both the APSN and Down modules clearly leads to noticeable style discrepancies between the generated sketches and hand-drawn sketches. Without the APSN module, there are evident problems with sketch lines and the handling of intricate details. Likewise, the absence of the Down module results in a loss of partial detail in the generated sketches. Overall, the results highlight the superiority of HAIGEN compared to the baselines and the contributions of individual modules in our model, showcasing its prowess in image-to-sketch generation. This makes a great contribution to generating a personalized sketch material library.

%%%%%%%%%%%%%%% Sketch Recommendation Module %%%%%%%%%%%%%%%%%%%
\subsection{Performance of Sketch Recommendation Module}
\subsubsection{Qualitative Results}

\begin{figure}[ht]
\begin{center}
\includegraphics[width=0.95\linewidth]{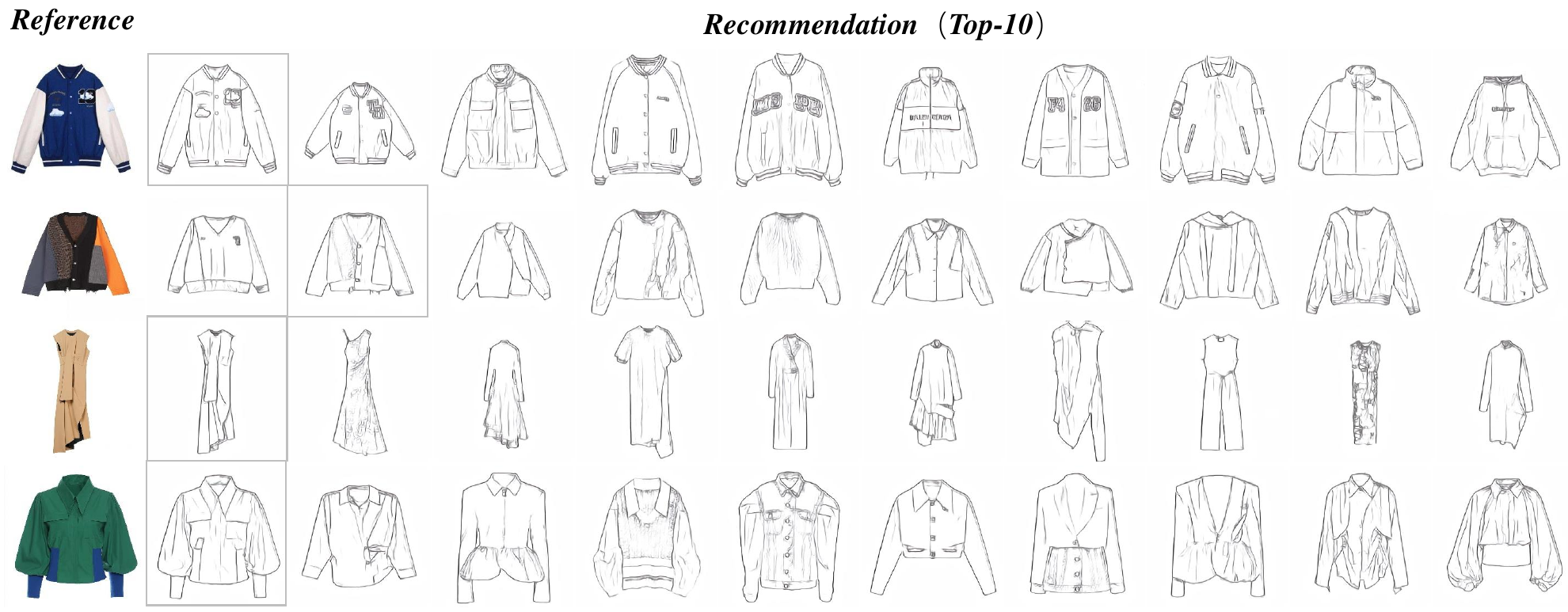}
\caption {The performance of our Sketch Recommendation Module on recommending similar sketch templates.}
\vspace{-3mm}
\label{fig:recom_gen}
\end{center}
\end{figure}

Fig.~\ref{fig:recom_gen} illustrates the impressive performance of HAIGEN in recommending sketch templates. The first column represents the input reference image, followed by ten columns showing the top-10 recommended sketch templates from our built sketch library, most closely resembling the reference image. Notably, the recommended sketches closely resemble the real sketches, appearing prominently in the lineup. Furthermore, other sketch templates also exhibit striking similarities with the reference images. For instance, similarities in clothing styles can be observed in the second and third rows, and details such as the neckline of the clothes and pockets in the first row are replicated in the recommended templates. Although the first-ranked recommended sketch template does not precisely match the reference image, the resemblance between this recommendation and both the reference image and the real sketch template is evident. These results convincingly highlight the exceptional performance of our Sketch Recommendation Module in recommending sketch templates

%%%%%%%%%%%%%%% Style Transfer Module %%%%%%%%%%%%%%%%%%%
\subsection{Performance of Style Transfer Module}
\subsubsection{Quantitative Results}
Table~\ref{table:s2i} illustrates the quantitative evaluation of our method using five metrics to compare the performance with other approaches. It is worth noting that our method exhibits clear advantages across all indicators, particularly excelling in LPIPS and FID, aligning with human eye perception and reaching new state-of-the-art levels. Specifically, our HAIGEN employs the DDIM algorithm with 100 samples, demonstrating substantial advantages. While the results with 20 and 50 samples exhibit slightly reduced accuracy, the benefits still stand out. For the 200 samples result, the increased time cost for interfacing doesn't provide a proportional improvement in image quality, making it less cost-effective. Furthermore, the comparison indicates that our proposed CCAM enhances the overall performance in all metrics. This improvement is attributed to its nature as an attention module, utilizing cross-attention for enhanced feature fusion between feature channels. The matrix operations in the other branch of CCAM facilitate simple aggregation of feature details. These quantitative results effectively demonstrate the effectiveness of our approach in sketch-to-image style transfer.

\begin{table}[tb]
\caption{Quantitative experimental results on sketch-to-image style transfer. ($\uparrow$ indicates the higher value, the better effect. $\downarrow$ indicates the lower value, the better effect.)}
\label{table:s2i}
\centering
\begin{tabular}{clccccc} \toprule
\multirow{2}{*}{} & \multirow{2}{*}{\textbf{Methods}} & \multicolumn{5}{c}{\textbf{HAIFashion}} \\
& & \textbf{PSNR} $\uparrow$ &\textbf{SSIM} $\uparrow$ &\textbf{MSE} $\downarrow$ &\textbf{LPIPS} $\downarrow$ &\textbf{FID} $\downarrow$ \\ \midrule \midrule
\multirow{5}{*}{\rotatebox[origin=c]{90}{Baselines}}
& \multicolumn{1}{l}{AdaIN~\cite{huang2017adain}} & 17.5611 & 0.3168 & 0.0302 & 0.1016 & 67.3568 \\
& \multicolumn{1}{l}{Self-Sup~\cite{liu2021self}} & 15.0941 & 0.3787 & 0.0446 & 0.1732 & 61.4071 \\
& \multicolumn{1}{l}{StyleMe~\cite{wu2023styleme}} & 22.8712 & \underline{0.4576} & 0.0099 & 0.0651 & 30.7081 \\
& \multicolumn{1}{l}{UCAST~\cite{zhang2023unified}} & 18.8464 & 0.1144 & 0.0230 & 0.1275 & 43.2199 \\
& \multicolumn{1}{l}{DiffusionRig~\cite{ding2023diffusionrig}} & 15.0146 & 0.1857 & 0.0359 & 0.1119 & 40.1118 \\ \midrule
\multirow{5}{*}{\rotatebox[origin=c]{90}{Ablation}}
& \multicolumn{1}{l}{w/o CCAM} & 22.1428 & 0.3625 & 0.0076 & 0.0363 & 19.8801 \\
& \multicolumn{1}{l}{20 samples} & 23.1550 & 0.2425 & 0.0058 & 0.0591 & 30.9403 \\
& \multicolumn{1}{l}{50 samples} & 22.9912 & 0.2748 & 0.0060 & 0.0350 & 19.9451 \\
& \multicolumn{1}{l}{200 samples} & \textbf{23.9765} & \textbf{0.5000} & $\textbf{0.0048}_7$ & \underline{$0.0334_4$} & \underline{18.1717} \\
& \multicolumn{1}{l}{\textbf{\textbf{HAIGEN (Ours)}}} & \underline{23.9430} & 0.3973 & \underline{$0.0049_3$} & $\textbf{0.0333}_7$ & \textbf{18.1367} \\ \bottomrule
\end{tabular}
\end{table}

\subsubsection{Qualitative Results}
\begin{figure}[ht]
\begin{center}
\includegraphics[width=0.95\linewidth]{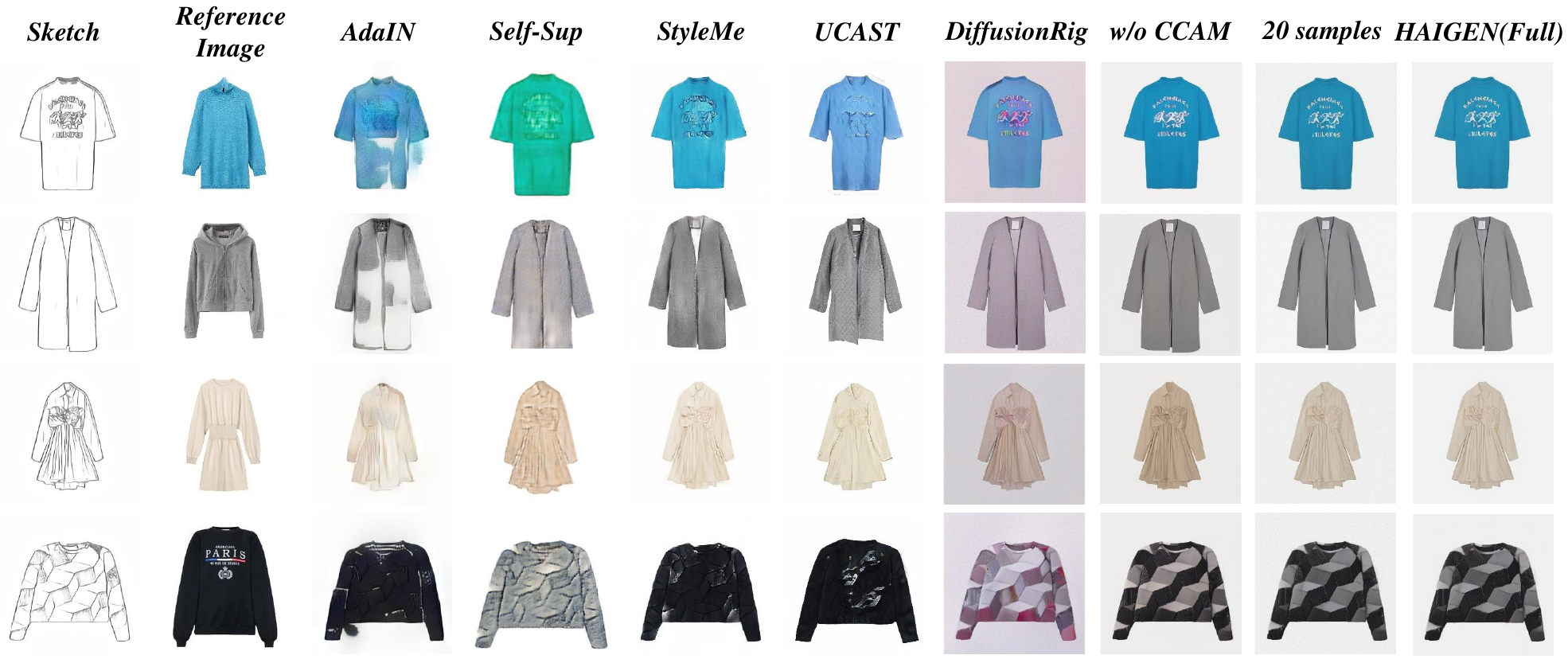}
\caption {The comparison of sketch-to-image synthesis from the baselines and our model.}
\vspace{-3mm}
\label{fig:s2i_gen}
\end{center}
\end{figure}

We demonstrate the image generation performance of our method compared with other baselines in sketch-to-image style transfer as shown in Fig.~\ref{fig:s2i_gen}. It's evident that the image quality generated by the diffusion model-based methods surpasses that of the GAN-based methods. For instance, DiffusionRig exhibits clear advantages in terms of generated image details, color uniformity, and color distribution, despite the obvious background color present in its images. The background color contributes to lower quantitative evaluation indicators. AdaIN faces challenges related to clothing integrity, color uniformity, and image detail. Self-Sup suffers from inconsistencies in color between the generated image and the reference image, along with a loss of details (first row) and an exaggeration of details (fourth row). StyleMe and UCAST, although exhibiting better generation effects than the former two, still struggle with retaining details. In comparison, our HAIGEN method demonstrates substantial advantages in pattern processing, detail retention, color distribution, and more. Notable examples include the patterns in the first row, clothing labels in the second row, texture details of the skirt in the third row, and patch color distribution in the fourth row. A further comparison with the CCAM removed reveals some lost details, darker colors, and inconsistent color matching with the reference. In summary, our method excels in maintaining a consistent color style while effectively handling the complete structure and intricate details of input clothing sketches. This robust performance underscores the effectiveness of our method in sketch coloring.

%%%%%%%%%%%%%%%%%%%%%%%%%%% User survey %%%%%%%%%%%%%%%%%%%%%%%%%%%
\section{User survey on HAIGEN system}
\label{sec:survey}
We further conducted user surveys to comprehensively assess the overall performance of our HAIGEN system and validate its effectiveness in inspiring design creativity and enhancing design efficiency.

\subsection{Experimental Settings}
We designed a set of controlled experiments, wherein one group utilized traditional clothing design methods. Designers searched for inspiration references through the Internet and progressed through the design process from ideas to sketches to images through hand drawing. In contrast, another group employed the HAIGEN system as an auxiliary tool. Designers use HAIGEN to generate inspiration references and facilitate the entire design process, as detailed in Fig.~\ref{fig:system}. Each group comprises three designers, all of whom are graduate students majoring in design and possess three years or more of design experience.
To avoid any potential skill disparities resulting from equipment replacement, all participating designers utilized their familiar equipment, including iPads and computers, during the experiments. And for the group utilizing the HAIGEN system, we configured the experimental environment on their existing equipment, provided preliminary training on the system's usage procedures, and ensured that all experiments were conducted on the same local area network to maintain consistency.

During the experiments, both groups of designers were assigned a specific design theme and tasked with creating clothing designs based on that theme. We conducted three separate experiments over three weeks, with each experiment assigned design themes of varying difficulty, including modern, retro, and tech. This approach was adopted to minimize experimental interference. Following the experiment, we gathered the six participating designers to share their feelings and experiences during the design process. Subsequently, we conducted in-depth interviews to gain further insights into their perspectives.

To assess the efficacy of the HAIGEN system in terms of the quality of completed works, we created a questionnaire regarding the design themes and the works they designed, which was distributed to the designer community. We collected 115 valid responses from the participants, collecting 115 valid responses. The questionnaire categorized participants' designs into three themes and further divided each theme into works created with or without the HAIGEN system. Participants used a Likert scale for evaluation, consisting of five ratings: "very good," "good," "average," "poor," and "very poor." This comprehensive approach allowed us to gather both qualitative and quantitative data on the impact of the HAIGEN system on design quality.

\subsection{Survey Results}
In the in-depth interviews conducted with the six participating designers, we gleaned the following insights: 
(i) Overall Performance: After using the HAIGEN system, designers obviously feel that designing is more convenient than before. 
(ii) Inspiration Stage: Designers can generate the desired inspiration images through detailed text descriptions rather than through complicated and repetitive searches. Moreover, it can switch between different scenes and styles while showing the clothing try-on renderings. In addition, the model can also generate many unexpected results for them, which makes them very excited.
(iii) Sketching Stage: The most satisfying thing for designers here is that they can modify the sketch directly on the recommended template instead of starting from scratch.
(iv) Coloring Stage: Designers find joy in the fact that they no longer need to visualize the coloring effects in their minds as required by traditional methods. Instead, they can directly employ the model to produce the corresponding renderings. This speeds up their coloring process and makes the coloring more accurate.

\begin{figure}[tb]
\begin{center}
\includegraphics[width=\linewidth]{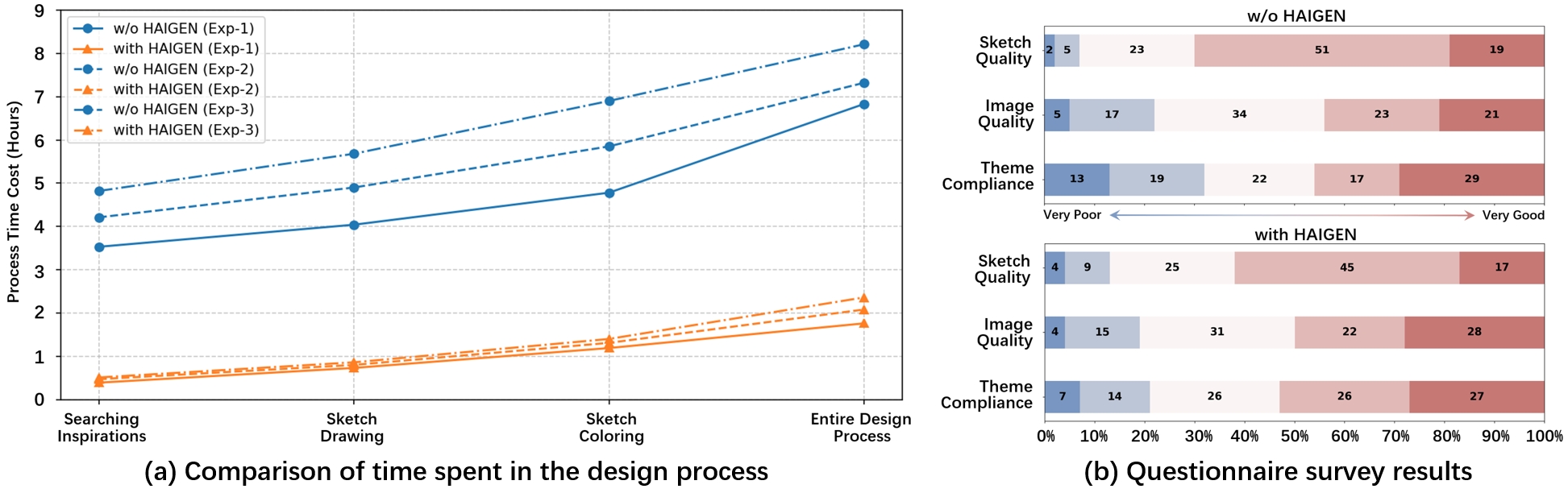}
\caption {The user survey results on time cost and work quality.}
\vspace{-3mm}
\label{fig:users}
\end{center}
\end{figure}

As depicted in Fig.~\ref{fig:users}, we present a comparison of the time spent in the design process when using the HAIGEN system (Fig.~\ref{fig:users} (a)), along with the results of the questionnaire survey regarding the obtained design works (Fig.~\ref{fig:users} (b)). The analysis reveals a consistent improvement in designers' efficiency across the three experimental groups after incorporating HAIGEN, with an impressive increase of approximately 250\%. This substantial enhancement is highly advantageous for their design processes. Notably, the traditional method requires around four hours for inspiration reference, whereas our approach achieves the same task in just about half an hour. Furthermore, in the sketching and sketch coloring stages, HAIGEN accelerates the design process by offering a large number of preset sketch templates and enabling designers to preview sketches in advance. The results of the questionnaire indicated a positive impact of HAIGEN on the overall quality of the final design work. Notably, it demonstrated commendable performance in terms of image quality and theme consistency, with the proportion of scores at "average" and above increasing by 3\% and 11\%, respectively. Additionally, positive reviews ("very good" and "good") increased from 44\% and 46\% to 50\% and 53\%, respectively. This success can be attributed to our system's capability to generate matching inspiration images from detailed text descriptions, providing rich background and clothing model try-on references, thereby effectively inspiring designers and preserving their initial ideas. Furthermore, the sketch coloring module enables designers to swiftly experiment with various styles, facilitating the identification of the most suitable color style for the final solution. Despite some negative feedback regarding sketch quality, attributed to limitations of the image-to-sketch generation algorithm in accurately generating line outlines outside of outlines like humans, the algorithm's fast generation speed and minimal impact of the sketches on the final image generation underscore its efficacy as a solution. Overall, these user experiments reaffirm the effectiveness of our system.

%%%%%%%%%%%%%%%%%%%%%%%%%%% Discussion %%%%%%%%%%%%%%%%%%%%%%%%%%%
\section{Discussion}
\label{sec:discussion}
\subsection{Industry Impact}
In our earlier user studies (Sec.~\ref{sec:study}), we discovered that designers now prefer using AI tools like Midjourney~\cite{midjourney} to enhance their design process, significantly improving their design efficiency. These AI tools enable designers to generate images based on detailed text descriptions, facilitating the effective expression of their creative ideas. Moreover, contributing to the integration of multiple modules, our HAIGEN system offers comprehensive functionality for the entire design process, benefiting both novice and professional designers alike. Designers can further enhance works generated by the HAIGEN system through secondary development using other design software or platforms, thereby achieving superior design effects.
Despite concerns regarding the emergence of AI, such as copyright and originality issues, and the potential for designers to become overly dependent on AI, we believe that judicious use of AI can offer numerous benefits to designers. Therefore, our focus in this paper is on leveraging AI as a tool to augment human creativity and design processes, fostering collaborative creation between humans and AI. While AI aids in enhancing efficiency and providing new ideas and inspiration, the ultimate decision-making and creative power remains in the hands of designers. Specifically, HAIGEN provides designers with inspiration references (Sec.~\ref{sec:sd}) or material templates (Sec.~\ref{sec:i2s}) and enables quick previews of coloring effects (Sec.~\ref{sec:s2i}) for their works, etc. Throughout these processes, designers play an active role in drawing and refining works based on AI-generated outputs. This is achieved through a strategic combination of deployment segregation and collaborative capabilities between HAIGEN's cloud large models and local small models. Such an approach effectively addresses concerns regarding design copyright infringement and leakage, while also mitigating the risk of excessive dependence on AI.

\subsection{Limitation and Future Work}
Although our system provides significant assistance to designers throughout the whole design process, there are still some weaknesses that need to be improved. Firstly, our system cannot currently mimic human-like thinking in generating sketches, instead relying on direct formation from the outline of images. Secondly, it cannot generate sketches of varying complexity tailored to the preferences of different designers. One potential solution for the first issue is to employ Bezier curves to simulate the drawing process of each stroke in the sketch, thereby achieving a more hand-drawn appearance. However, this approach introduces new challenges, particularly in the coloring process. Generating concrete images from abstract sketches poses a significant challenge as there is no strict one-to-one correspondence between the strokes of the sketch generated using Bezier curves and the corresponding image contours. This remains a problem that current algorithms struggle to solve effectively. As for the second issue, a feasible solution is to introduce a separate hyperparameter to control the complexity of generated sketches. This would allow designers to tailor the output to meet their specific design needs and facilitate targeted refinement. In summary, our future research will focus on exploring more general and applicable AI-assisted design solutions to address these challenges and further enhance the capabilities of our system.

%%%%%%%%%%%%%%%%%%%%%%%%%%% Conclusions %%%%%%%%%%%%%%%%%%%%%%%%%%%
\section{Conclusions}
\label{sec:conclusion}

In this paper, we introduce HAIGEN, a Human-AI collaboration fashion design system that seamlessly integrates designers with cloud large models and local small models. This integration facilitates the entire design creation process, addressing the specific needs identified through user study. First and foremost, we developed an enhanced cloud-based Stable Diffusion large model that can generate a wealth of relevant inspirational images, serving as a wellspring of creative ideas for designers. Moreover, we've integrated a sketch generation model renowned for capturing designers' unique styles. Coupled with our self-collected fashion clothing image material library, this model can generate a profusion of personalized sketch templates. To further streamline the sketching process, we've introduced a sketch recommend module that recommends similar sketch templates to designers. This approach simplifies and expedites the sketch creation process. In the final phase, our style transfer module, equipped with impressive feature fusion capabilities, harmoniously blends refined sketches with inspirational images. The outcome is a collection of high-quality, realistic, and expertly-colored sketch images. To validate the performance of our system, we conducted extensive qualitative and quantitative experiments on our self-collected clothing datasets. The results underscore the efficacy and utility of each module within our system. Furthermore, through user usage surveys, we emphasize the applicability and effectiveness of HAIGEN, establishing it as an efficient generative AI system for fashion design with human-in-the-loop.

% generative AI for fashion design with human-in-the-loop

% \clearpage

%%
%% The next two lines define the bibliography style to be used, and
%% the bibliography file.
\bibliographystyle{ACM-Reference-Format}
\bibliography{main}

%%
%% If your work has an appendix, this is the place to put it.
\appendix

\end{document}